\documentclass[11pt]{article}
\usepackage[utf8]{inputenc}
\usepackage{amsmath}
\usepackage[dvipsnames]{xcolor}
\usepackage{amsthm}
\usepackage{pdflscape}
\usepackage{pgfplots}
\pgfplotsset{compat=1.18}
\usepackage{mathrsfs}
\usepackage{tikz}
\usepackage{amssymb}
\usepackage{forest}
\usepackage[hmargin=1.25in]{geometry} 
\usepackage{graphicx}
\usepackage{float}
\usepackage[colorlinks,citecolor=teal,linkcolor=blue,urlcolor=black,pagebackref]{hyperref}
\usepackage[capitalise]{cleveref}
\usepackage{enumerate}
\usepackage[shortlabels]{enumitem}
\usepackage{algorithm}
\usepackage{algpseudocode}
\usepackage{thmtools}
\usepackage{thm-restate}
\usepackage{titlesec}
\usepackage{alphalph}
\usepackage{comment}

\usepackage{caption}
\usepackage[indention=20pt]{subcaption}

\makeatletter
\DeclareRobustCommand{\cev}[1]{%
  {\mathpalette\do@cev{#1}}%
}
\newcommand{\do@cev}[2]{%
  \vbox{\offinterlineskip
    \sbox\z@{$\m@th#1 x$}%
    \ialign{##\cr
      \hidewidth\reflectbox{$\m@th#1\vec{}\mkern4mu$}\hidewidth\cr
      \noalign{\kern-\ht\z@}
      $\m@th#1#2$\cr
    }%
  }%
}
\makeatother

\newcommand{\cD}{\mathcal{D}}
\newcommand{\cC}{\mathcal{C}}
\newcommand{\cP}{\mathcal{P}}

\newcommand{\Z}{\mathbb{Z}}

\newcommand{\N}{\mathbb{N}}

\newcommand{\qacc}{q_{\textup{accept}}}
\newcommand{\qrej}{q_{\textup{reject}}}
\newcommand{\cart}{\, \square \,}

\theoremstyle{plain}
\newtheorem{theorem}{Theorem}[section]
\newtheorem{corollary}[theorem]{Corollary}
\newtheorem{proposition}[theorem]{Proposition}
\newtheorem{lemma}[theorem]{Lemma}
\newtheorem{claim}[theorem]{Claim}

\theoremstyle{definition}
\newtheorem{definition}[theorem]{Definition}
\newtheorem{assumption}[theorem]{Assumption}

\theoremstyle{remark}
\newtheorem{remark}[theorem]{Remark}

\newcommand{\countH}{{\bf \#H}}

\newcommand{\set}[1]{\left\{#1\right\}}

\newcommand{\poly}{{\rm poly}}
\newcommand{\bigO}[1]{O\left(#1\right)}

\newcommand{\nspace}{{\bf NSPACE}}

\newcommand{\dtime}{{\bf DTIME}}

\newcommand{\xp}{\mathbf{XP}}

\newcommand{\xnl}{\mathbf{XNL}}
\newcommand{\fpt}{\mathbf{FPT}}
\newcommand{\ptime}{{\bf P}}
\newcommand{\nl}{{\bf NL}}

\newcommand{\tiles}{\hat{Q} \times \hat{\Gamma}}
\newcommand{\state}{\mathrm{state}}
\newcommand{\alp}{\mathrm{alph}}
\newcommand{\cont}{\mathrm{cont}}
\newcommand{\lfmis}{\mathrm{LFMIS}}

\newcommand{\actS}{{\rm ACTIVE}}
\newcommand{\inactS}{{\rm INACTIVE}}
\newcommand{\nullS}{\bot_{S}}
\newcommand{\nullI}{\bot_{I}}
\newcommand{\nullW}{\bot_{W}}
\newcommand{\sconfig}{{\cal F}}

\title{The Computational Complexity of Factored Graphs}

\author{Shreya Gupta\thanks{University of California, San Diego. \href{mailto:sfgupta@ucsd.edu}{sfgupta@ucsd.edu}.},
Boyang Huang\thanks{University of California, San Diego. \href{mailto:boh002@ucsd.edu}{boh002@ucsd.edu}.},
Russell Impagliazzo\thanks{University of California, San Diego. \href{mailto:rimpagliazzo@ucsd.edu}{rimpagliazzo@ucsd.edu}. Supported by NSF Award AF: Medium 2212136.}, 
Stanley Woo\thanks{University of California, San Diego. \href{mailto:tlwoo@ucsd.edu}{tlwoo@ucsd.edu}.},
Christopher Ye\thanks{University of California, San Diego. \href{mailto:czye@ucsd.edu}{czye@ucsd.edu}. Supported by NSF Award AF: Medium 2212136, NSF grants 1652303, 1909046, 2112533, and HDR TRIPODS Phase II grant 2217058.}}

\begin{document}

\maketitle
\pagenumbering{gobble}

\begin{abstract}
    While graphs and abstract data structures can be large and complex, practical instances are often regular or highly structured. If the instance has sufficient structure, we might hope to compress the object into a more succinct representation. An efficient algorithm (with respect to the compressed input size) could then lead to more efficient computations than algorithms taking the explicit, uncompressed object as input. This leads to a natural question: when does knowing the input instance has a more succinct representation make computation easier?

We initiate the study of the computational complexity of problems on factored graphs: graphs that are given as a formula of products and unions on smaller graphs. For any graph problem, we define a parameterized version that takes factored graphs as input, parameterized by the number of (smaller) ordinary graphs used to construct the factored graph.
In this setting, we characterize the parameterized complexity of several natural graph problems, exhibiting a variety of complexities.
We show that a decision version of lexicographically first maximal independent set is \textbf{XP}-complete, and therefore \emph{unconditionally} not fixed-parameter tractable (\textbf{FPT}). On the other hand, we show that clique counting is \textbf{FPT}. Finally, we show that reachability is \textbf{XNL}-complete. Moreover, \textbf{XNL} is contained in \textbf{FPT} if and only if \textbf{NL} is contained in some fixed polynomial time.
\end{abstract}
\newpage

\tableofcontents
\newpage

\pagenumbering{arabic}
\setcounter{page}{1}

\section{Introduction}
\label{sec:introduction}

Traditionally, algorithm design and computational complexity both measure computational time as a function of the input size.
Thus, the complexity of computational problems is crucially sensitive to the way the instances are represented as bit sequences.
While graphs and abstract data structures can be complex and expressive in the worst case, practical instances are often highly structured or regular. 
For example, road networks are often organized into repetitive grid patterns. Similarly, in databases, relations frequently inherit underlying structures from previous relations through operations like joins. In molecular geometry, compounds such as graphite are composed of layers of graphene, with each layer forming a regular honeycomb structure. If the input instance has sufficiently regular structure, we might hope to compress the object into a more succinct representation.

An algorithm that is efficient with respect to the size of the succinct encodings would then be able to compute the desired result more efficiently than an algorithm that takes a naive representation of the input.
This raises the question: when does knowing that the input instance was created in a uniform way (or has a succinct representation) make computation easier?

To address this question, various formulations of ``succinct representation'' have been explored. For example, one such formulation is given by a Boolean circuit that can produce any particular input bit \cite{galperin1983succinct, viola2018localexpanders}.
Frequently, the complexity of problems given in this type of succinct format is exponentially more difficult than when the input is given explicitly (see e.g.\ \cite{galperin1983succinct, feigenbaum1995game, marathe1994approximation, marathe1998theory}). However, it's important to note that the difficulty only increases as a function of the smaller size parameter, and the actual problem has not necessarily become more difficult. 

Another type of succinct representation, factored instances, was introduced by \cite{dalirrooyfard2020factor}.
While they give a number of specific problems rather than introduce the concept abstractly, we can generalize by thinking of a fixed set of operations that take pairs of instances to possibly larger instances.  
For example, one natural operation might be set sum, taking two sets of integers $A, B$ to the set of all sums $A + B = \set{a + b \ |\ a \in A, b \in B}$.
Instead of being described directly, the input to a factored problem is given as this operation applied to a pair of possibly smaller instances. 
If these operations can be computed in polynomial time, the size of the output might be polynomially larger than the input, so this representation might be considerably smaller than the original. 
For each such set of operations and problem on instances, we can define a parameterized version of the problem on factored instances, where the input is represented as a formula in these (binary) operations over smaller instances, with the parameter $k$ being the number of smaller instances.  
If the underlying problem is polynomial time solvable, then for fixed $k$, the problem remains polynomial time solvable, but with an exponent that potentially grows with $k$.
How the compressed representation affects the difficulty of the problems can be formalized in terms of the complexity of such factored problems from the point of view of parameterized complexity. 
In this paper, we consider such parameterized factored problems for graphs, using standard graph products and union as our operations.

To make this precise, we first review some standard concepts from parameterized complexity. 
The gold standard for tractability  in parameterized complexity is membership in the class of fixed-parameter tractable ($\fpt$) problems \cite{downey1995fixed}.
Roughly speaking, a factored problem will be fixed-parameter tractable (i.e. in $\fpt$) if on an instance composed using $k$ smaller instances of size at most $n$, there is an algorithm computing the function in $O(f(k) n^{C})$ time for some fixed function $f$ and absolute constant $C$ independent of $k$.
In these settings, we think of $n$ as large and $k$ as small, so that any dependence on $k$ alone is preferable over an exponent of $n$ that grows with $k$.
A natural approach to solving problems on factored graphs is to explicitly compute the graph, requiring time and space $n^{O(k)}$, and solving the problem on the explicit graph.
Such an algorithm has an exponent growing with $k$.  
Thus such factored problems will always be in $\xp$\footnote{assuming the problem can be solved in polynomial time on the explicit graph}, the class of problems that are polynomial for any particular $k$, but not necessarily in $\fpt$.
Even for simple problems with linear time algorithms (on explicit graphs), computing an explicit representation already requires $n^{O(k)}$ time and so this approach does not put the problem in $\fpt$.

In this work, we initiate the study of the complexity of computational problems on factored graphs.
In particular, we consider the question: for a given computational problem, is there an algorithm substantially better than first explicitly computing the factored graph $G$?
This can be formalized as: is the problem in $\fpt$, and, if not, how does the best exponent of $n$ possible depend on $k$?

\subsection{Definition of Factored Graphs}

We consider a highly natural class of factored problems where the instances are graphs, and the operations are well-studied graph products and unions.
Our factored graphs consist of arbitrary combinations of graphs under the following (binary) operations.
Unless otherwise stated, all graphs are directed and $(a, b)$ denotes an edge from $a$ to $b$.
 
\begin{enumerate}
    \item {\bf Cartesian Product.} Given $A, B$, the Cartesian product $A \cart B$ has vertices $(a, b)$ for $a \in A, b \in B$ and there is an edge from $(a, b)$ to $(c, d)$ if either 1) $a = c$ and $(b, d) \in E(B)$, or 2) $(a, c) \in E(A)$ and $b = d$.
    \item {\bf Tensor Product.} Given $A, B$, the tensor product $A \times B$ has vertices $(a, b)$ for $a \in A, b \in B$ and there is an edge from $(a, b)$ to $(c, d)$ if $(a, c) \in E(A)$ and $(b, d) \in E(B)$.
    \item {\bf Union.} Given $A, B$, the union $A \cup B$ has vertices $x \in V(A) \cup V(B)$ and there is an edge from $x$ to $y$ if $(x, y) \in E(A) \cup E(B)$.
\end{enumerate}

A factored graph $G$ can be defined by an ordered tree $T(G)$, where internal nodes are labelled by one of the above three operations $\square, \times, \cup$ and leaves are labelled with (possibly identical) input graphs.
Since each operation is associative, we in fact allow the internal nodes to have arbitrary degree.\footnote{We require each internal node to have degree greater than 1, as otherwise the operation is the identity.}
We say that a factored graph $G = f(G_1, \ldots, G_k)$ is \emph{of complexity $(n, k)$} if the tree has $k$ leaves and each input graph has at most $n$ vertices. 
Here, we stress that the we parameterize by the number of leaves in $T(G)$, regardless of the fan-in of specific internal nodes or if a certain input graph appears in multiple leaf nodes. As an example, both $(A \times B) \times A$ and $A \times B \times A$ are factored graphs of complexity $(n, 3)$, where $n = \max(|V(A)|, |V(B)|)$.
We illustrate their tree structures in \Cref{fig:simple-factored-graphs}.

\begin{figure}[ht]
    \centering

    \tikzstyle{node}=[circle, draw, minimum size=10mm, inner sep=0pt]
    \usetikzlibrary{shapes.geometric}

    \begin{minipage}[t]{0.45\textwidth}
        \centering
        \begin{tikzpicture}[thick, level distance=1.25cm, level 1/.style={sibling distance=3cm}, level 2/.style={sibling distance=1.5cm}]
        \node[node] {$\times$}
            child {node[node] {$\times$}
                child {node[node] {$A$}}
                child {node[node] {$B$}}
            }
            child {node[node] {$A$}};
    \end{tikzpicture}
        \caption*{Tree structure of $(A \times B) \times A$.}

    \end{minipage}
    \hfill
    \begin{minipage}[t]{0.45\textwidth}
        \centering
        \begin{tikzpicture}[thick, level distance=2cm, level 1/.style={sibling distance=1.5cm}]
        \node[node] {$\times$}
            child {node[node] {$A$}}
            child {node[node] {$B$}}
            child {node[node] {$A$}};
        \end{tikzpicture}
        \caption*{Tree structure of $A \times B \times A$.}
        
    \end{minipage}

\caption{Tree structures of factored graphs $(A \times B) \times A$ and $A \times B \times A$. Both factored graphs have complexity $(n, 3)$ where $n = \max(|V(A)|, |V(B)|)$. Note that in this example the two factored graphs are in fact isomorphic.}
\label{fig:simple-factored-graphs}
\end{figure}

As another example, any graph $G$ can be represented by a factored graph of complexity $(2, |E(G)|)$ by taking a union over all edges of the graph.

We define a \emph{factored component} of $G$, denoted $G_{F}$, as follows.
$G_{F}$ is defined by an ordered tree $T(G_{F})$, which is obtained from $T(G)$ by deleting every internal node $v$ labeled by $\cup$ and attaching exactly one child of $v$ to its parent.
Another way to understand this is that, since the union and product operations obey the distributive law, we can recursively apply this law to convert a factored graph formula into a union of products.
Then, each of these product terms is a factored component.
The vertices of $G_{F}$ are $k_{F}$-tuples (note $k_{F} \leq k$, where $k_F$ is the number of leaves of $T(G_F)$ and $k$ is the number of leaves of $T(G)$).
We emphasize that the vertices are \emph{flattened} tuples and do not preserve the topology of $T(G_{F})$ beyond the order of the leaves.
We call $k_{F}$ the \emph{dimension} of the factored component $G_{F}$.
The edges of $G_{F}$ are determined by $T(G_{F})$ according to the definition of the graph operations $\cart$ and $\times$.

The vertex and edge sets of $G$ are then given by the union of the vertex and edge sets of the factored components.
While a vertex may belong to multiple factored components, they must all have the same dimension, allowing us to define the dimension of a vertex (\Cref{def:vertex-dimension}).
We also observe that $G$ has at most $2^{k}$ factored components.
We give a more formal definition of factored graphs as well as simple examples in \Cref{sec:prelims}.

\subsection{Our Contributions}

Our first result provides an {\em unconditional} hardness for a natural parameterized version of the well-studied Lexicographically First Maximal Independent Set (LFMIS) problem.
The standard decision version of the LFMIS problem takes as input a graph $G = (V(G), E(G))$, where the vertices are indexed as $V(G) = \{0, 1, \dotsc, |V(G)| - 1\}$, and a special vertex $s \in V(G)$.
The problem asks whether $s$ belongs to the LFMIS of $G$. 
In the parameterized version, the input is given as a factored graph $G = f(G_1, \dotsc, G_k)$ and the vertex indices are provided only for each graph factor $V(G_i) = \{0, 1, \dotsc, |V(G_i)| - 1\}$.
To recover the indices for $V(G)$, recall that each vertex in the factored graph $G$ is a flattened tuple of numbers $(v_1, \dotsc, v_{k'})$ for some $k' \leq k$, and therefore, we define the vertex indices to be given according to the standard lexicographic order of these tuples, with the index 0 given to the vertex with the lowest lexicographic order.
In this work, we show that the LFMIS problem on factored graphs is $\xp$-complete under $\fpt$-reductions\footnote{See Definition 9.3 \cite{downey2012parameterized}. Also formally stated in \Cref{def:xp-xnl-complete} in the context of factored graph problems.} and therefore \emph{unconditionally} not in $\fpt$, providing a lower bound on the best possible exponent as a function of $k$.
LFMIS is a natural $\ptime$-complete problem \cite{cook1985taxonomy}, studied by \cite{uehara1997measure, uehara1999another, uehara1999measure} including in the parallel setting \cite{coppersmith1987parallel, miyano1989lexicographically, calkin1992expected, blelloch2012greedy}.  
We use the $\ptime$-completeness of this problem as an intuition for why the factored version might be difficult ($\xp$-complete), but we know of no direct connection between $\ptime$-completeness and hardness of the factored version.  

\begin{restatable}[$\xp$-completeness of LFMIS]{theorem}{LFMISXPC}
    \label{thm:factored-LFMIS-XP-complete}
    The LFMIS problem on factored graphs is $\xp$-complete under $\fpt$-reductions and not fixed-parameter tractable. 
    In particular, the LFMIS problem on a factored graph $G = f(G_1, \ldots, G_k)$ of complexity $(n, k)$ requires $n^{\Omega (\sqrt{k})}$ time.
\end{restatable}

While many works in parameterized complexity gave \emph{conditional} lower bounds against problems in $\fpt$ \cite{downey1995fixed}, for example completeness for the \textbf{W}-hierarchy \cite{downey1995fixed, downey1995fixed2, downey1996descriptive} or the \textbf{A}-hierarchy \cite{flum2001fixed}, \emph{unconditional} lower bounds against natural problems in $\fpt$ are relatively rare \cite{downey2012parameterized}.
Moreover, although it is known that $\fpt \subsetneq \xp$ via a diagonalization argument, the literature on $\xp$-complete problems remains sparse. 
The ``pebble game'' problem was first introduced by Kasai, Adachi, and Iwata \cite{kasai1979pebblegames}.
Its parameterized version, the ``$k$-pebble game'', was one of the earliest natural combinatorial problems shown to require $\Omega(n^k)$ time and be $\xp$-complete \cite{adachi1984combinatorialgame}.
As an application of the $k$-pebble game, some other game problems, such as the ``cat and mouse game'' \cite{chandra1976alternation} and the ``$k$-peg game'', have also been proven to be $\xp$-complete via reductions from the $k$-pebble game problem \cite{adachi1984combinatorialgame}.
More recently, Berkholz established an unconditional lower bound of $O(n^{(k-3)/12})$-time for the existential $k$-pebble game \cite{berkholz2012k-existential-pebble-game}.
Our result provides yet another natural example of an $\xp$-complete problem with an explicit lower bound of $n^{\Omega({\sqrt{k}})}$.

In contrast to the LFMIS problem on factored graphs, we show that the well-studied problem of clique counting \cite{nevsetvril1985complexity} is in $\fpt$, solvable in a fixed polynomial of the input graph sizes (where the exponent possibly depends on the size of the clique of interest) times some function on the number of input graphs. 
Subgraph counting (in particular clique counting) has been studied by a long line of works \cite{nevsetvril1985complexity, chen2006strong, calabro2008complexity, jain2020power} in a variety of computational models \cite{dalirrooyfard2020factor, dhulipala2021parallel, boix2021average, shi2021parallel, henzinger2022complexity}.
On graphs with $n$ vertices, $s$-cliques can be counted in $O(n^{s})$ time (and $n^{\Omega(s)}$ time is necessary under standard hardness assumptions). 
We show that counting $s$-cliques on factored graphs of complexity $(n, k)$ is in $O(g(s, k) n^{s})$ time for some fixed (large) function $g$.
Note that we do not show clique counting is in $\fpt$ with respect to the clique size parameter $s$, but only with respect to the complexity of the factored graph $k$.

\begin{restatable}[Counting Clique Subgraphs is in $\fpt$]{theorem}{CountingSubgraphs}
    \label{thm:counting-subgraphs}
    Let $H$ be a clique on $s$ vertices.
    Then, computing the number of exact copies of $H$ in $G$, denoted $\countH(G)$, is fixed-parameter tractable.
    In particular, there is an algorithm computing $\countH(G)$ on factored graphs $G$ of complexity $(n, k)$ in $O\left( g(s, k) n^{s} \right)$ time for some function $g(s, k)$.
\end{restatable}

Finally, we turn to the problem of reachability, one of the most fundamental computational problems on graphs.
On an explicit graph, algorithms such as depth-first search (DFS) and breadth-first search (BFS) compute reachability in linear time.
This raises the question: can reachability on factored graphs be computed efficiently?
Moreover, reachability is closely related to classic space complexity classes. 
The general reachability problem is an important $\nl$-complete problem \cite{sipser1996introduction, jones1975spacebounded}; meanwhile, Cook and Mckenzie showed that reachability on a directed acyclic graph of outdegree at most one is $\mathbf{L}$-complete \cite{cook1987logspacecomplete}.
This naturally gives rise to another question: does the completeness of reachability for a classic complexity class extend to its parameterized counterpart when the inputs are given as factored graphs, similar to what we have seen in the case of LFMIS in \Cref{thm:factored-LFMIS-XP-complete}?

Indeed, we show that the parameterized version of reachability on factored graphs is $\xnl$-complete under $\fpt$-reductions, where $\xnl$ is the class of parameterized problems solvable in nondeterministic logarithmic space for any fixed parameter $k$.
On the other hand, while we cannot provide a definitive answer to whether reachability on factored graphs is in $\fpt$, we show that answering this question would exactly resolve a major open problem in classical complexity theory.
Specifically, we establish a parameterized analog of the well-known open problem that asks whether $\nl \subseteq \dtime(n^{C})$ for some fixed constant $C$.

\begin{restatable}[$\xnl$-completeness of Reachability]{theorem}{ReachabilityXNLC}
    \label{thm:reachability-xnl-complete}
    Reachability on factored graphs is $\xnl$-complete under $\fpt$-reductions.
    Furthermore, the following are equivalent:
    \begin{enumerate}
        \item There is a constant $C$ such that $\nl \subseteq \dtime(n^{C})$.
        \item $\xnl \subseteq \fpt$.
    \end{enumerate}
    
\end{restatable}

This result also has several further implications.
First, we observe that if reachability is not in $\fpt$, $\nspace(h(n)) \not\subseteq \ptime$ for any space-constructible $h(n) = \omega(\log n)$.
On the other hand, if reachability is in $\fpt$ then the Exponential Time Hypothesis (ETH) is false. 
In particular, we claim that since $k$-SUM (determining if in a set of $n$ numbers of $O(k \log n)$ bits there is a subset of $k$ numbers summing to zero) is in $\nl$, then $k$-SUM can be solved in $O\left(n^{C}\right) = n^{o(k)}$ time, and therefore ETH is false \cite{puatracscu2010possibility}.\footnote{We need $O(k^2 \log n)$ bits to verify that $k$ numbers of $O(k \log n)$ bits sum to zero.}
Finally, by combining the above implications, one can also conclude that if ETH is true, then $\nspace(\omega(\log n)) \subsetneq \ptime$, a result that, to the best of our knowledge, has not been previously established.

Parameterized space complexity classes, including \textbf{XL} and \textbf{XNL}, have also been studied extensively in the literature \cite{cai1997advice-class-param-tractability, chen2003bounded-nondeterminism-param-complexity, flum2003param-complexity-classes, elberfeld2012space-complexity-param}.
Chen, Flum, and Grohe introduced the first complete problems for \textbf{XL} and \textbf{XNL} under parameterized-logspace-reductions\footnote{Also known as \textbf{PL}-reductions, a more restrictive notion compared to $\fpt$-reductions} \cite{chen2003bounded-nondeterminism-param-complexity}.
In a related direction, \cite{bodlaender2023parameterized} identified a wide variety of problems complete for the class \textbf{XNLP}, which is the class of parameterized problems solvable in nondeterministic logarithmic space \emph{and} polynomial time for each parameter.
Most relevant to our result, Elberfeld, Stockhusen, and Tantau also showed that a parameterized version of reachability with multi-colored edges is complete for the class parameterized-\textbf{NL}-cert under \textbf{PL}-reductions \cite{elberfeld2012space-complexity-param}.

\subsection{Technical Overview}

We now give a technical overview of our results and proof techniques.

\subsubsection{Lexicographically First Maximal Independent Set}\label{sec:LFMIS-tech-overview}

\Cref{thm:factored-LFMIS-XP-complete} is a consequence of a generic reduction from an arbitrary language $L \in \dtime(n^{\ell})$ to the LFMIS problem on factored graphs.
Specifically, given a language $L \in \dtime(n^{\ell})$ and the corresponding Turing machine that decides $L$ in $O(n^{\ell})$ time, we construct a factored graph $G$ of complexity $(O(n), O(\ell^2))$ using $O(\ell^2 n^2)$ time, such that solving the LFMIS on $G$ simulates the Turing machine $M$.
Then, it is fairly straightforward from this reduction that the problem is $\xp$-complete.
Moreover, we also obtain the explicit lower bound of $n^{\Omega(\sqrt{k})}$ by an application of the Time Hierarchy Theorem to the reduction.

Before proceeding to the detailed explanation, we first provide a high level intuition of this construction.
Given a Turing machine, we can encapsulate its computation history in a matrix, where each row represents a configuration of the Turing machine at a specific time. 
The key idea of the reduction is to construct a graph in such a way that selecting vertices for the LFMIS corresponds to recovering the matrix entries, and thereby simulating the machine.
To achieve this, we construct a graph with a grid structure mirroring the matrix.
At each grid point, we place a collection of vertices representing all the possible choices for the corresponding matrix entry.
The edges are defined according to the machine's transition function to ensure that the LFMIS chooses the single correct vertex from each grid point that agrees with the computation history of the Turing Machine, effectively allowing us to simulate the Turing machine by solving the LFMIS on the graph. 
It turns out that this graph has a highly regular structure and therefore can be factorized into a more succinct representation.
In the remainder of this (sub)section, we explain this reduction in further detail.

Let $L$ be a language in $\dtime(n^{\ell})$ and let $M$ be the corresponding Turing machine that decides $L$ in time $O(n^{\ell}).$
Given an input $x$ to $M$, if $M$ halts within time $T$, then the entire computation history of $M$ on $x$ can be represented by a $T \times T$ matrix $W$, where the $i$-th row of $W$ represents the configuration of $M$ at time $i$.
To achieve this, each entry $W_{i, j}$ contains the following information about the $j$-th tape cell at time $i$: 1) the symbol occupying the cell, 2) whether the tape head is over the cell, and 3) the machine's current state if the tape head is over the cell.
The goal now is to define a graph $G$ such that the LFMIS of $G$ recovers the computation history $W$, thereby simulating the machine $M$.
We modify the Turing machine so that it suffices to query a single vertex in the graph for the halting state of $M$ (see \Cref{sec:Turing-machine-model} for full details on the Turing machine model).

\paragraph*{Explicit Graph Definition}
We begin by defining $G$ explicitly as an ordinary graph and then give a factored representation of $G$.
Let $S$ be the set of vertices corresponding to all possible choices for an entry in the matrix $W$.
We define the vertex set of $G$ to be $T^2$ copies of $S$, arranged in a $T \times T$ grid-like pattern analogous to the matrix $W$.
Each copy of $S$ is referred to as a \emph{supernode}, and we use $S_{i, j}$ to denote the supernode in the $i$-th row and $j$-th column.
Intuitively, the supernode $S_{i, j}$ represents all possible choices for the entry $W_{i, j}$.
Therefore, we must define the edges of $G$ such that the LFMIS includes the single correct vertex, which we call $w_{i, j}$, from each $S_{i, j}$ that agrees with the entry $W_{i, j}$.

The edge set of $G$ consists of two types of edges: \emph{intra}-supernode and \emph{inter}-supernode edges.
The intra-supernode edges are designed to ensure that the LFMIS contains \emph{at most one} vertex within each supernode.
This is easily achieved by defining a complete digraph on each supernode.
On the other hand, the inter-supernode edges are constructed to ensure that the LFMIS contains \emph{the correct} vertex from each supernode that agrees with the corresponding entry of $W$.
This construction is more subtle and leverages the fact that the configuration at any given time of a deterministic machine uniquely determines the next configuration.
Moreover, since the tape head can only move one step left or right at a time, each entry $W_{i, j}$ can be uniquely determined given only the three neighboring entries from the previous row: $W_{i-1, j-1}, W_{i-1, j}, W_{i-1, j+1}$ (instead of the entire previous row). 
To view this from another perspective: the entries $W_{i-1, j-1}$, $W_{i-1, j}$, and $W_{i-1, j+1}$ each restricts the possible choices for $W_{i, j}$ according to the machine's transition function.
We define edges between neighboring supernodes to encode these restrictions. 
Specifically, for $v \in S_{i, j}$ and $v' \in S_{i + 1, j'}$, a directed edge $(v, v')$ indicates that, if $v$ is chosen to represent the $j$-th tape cell at time $i$, then $v'$ cannot represent the $j'$-th tape cell at the next time step $i + 1$, based on the machine's transition function.
We hope that if the vertices $w_{i - 1, j - 1} \in S_{i - 1, j - 1}$, $w_{i-1, j} \in S_{i-1, j}$, and $w_{i-1, j+1} \in S_{i-1,j+1}$ are correctly contained in the LFMIS, then the edges ensure that $w_{i, j}$ is the unique vertex in $S_{i, j}$ that is not adjacent to $w_{i - 1, j - 1}, w_{i - 1, j},$ or $w_{i - 1, j + 1}$.
This will allow us to build the LFMIS inductively.
The base case (which corresponds to the first row) is more technical and involves adding horizontal connections between supernodes.
We leave the detailed discussion to \Cref{sec:LFMIS}.

\begin{figure}[ht]
    \begin{center}
        \includegraphics[scale=0.3]{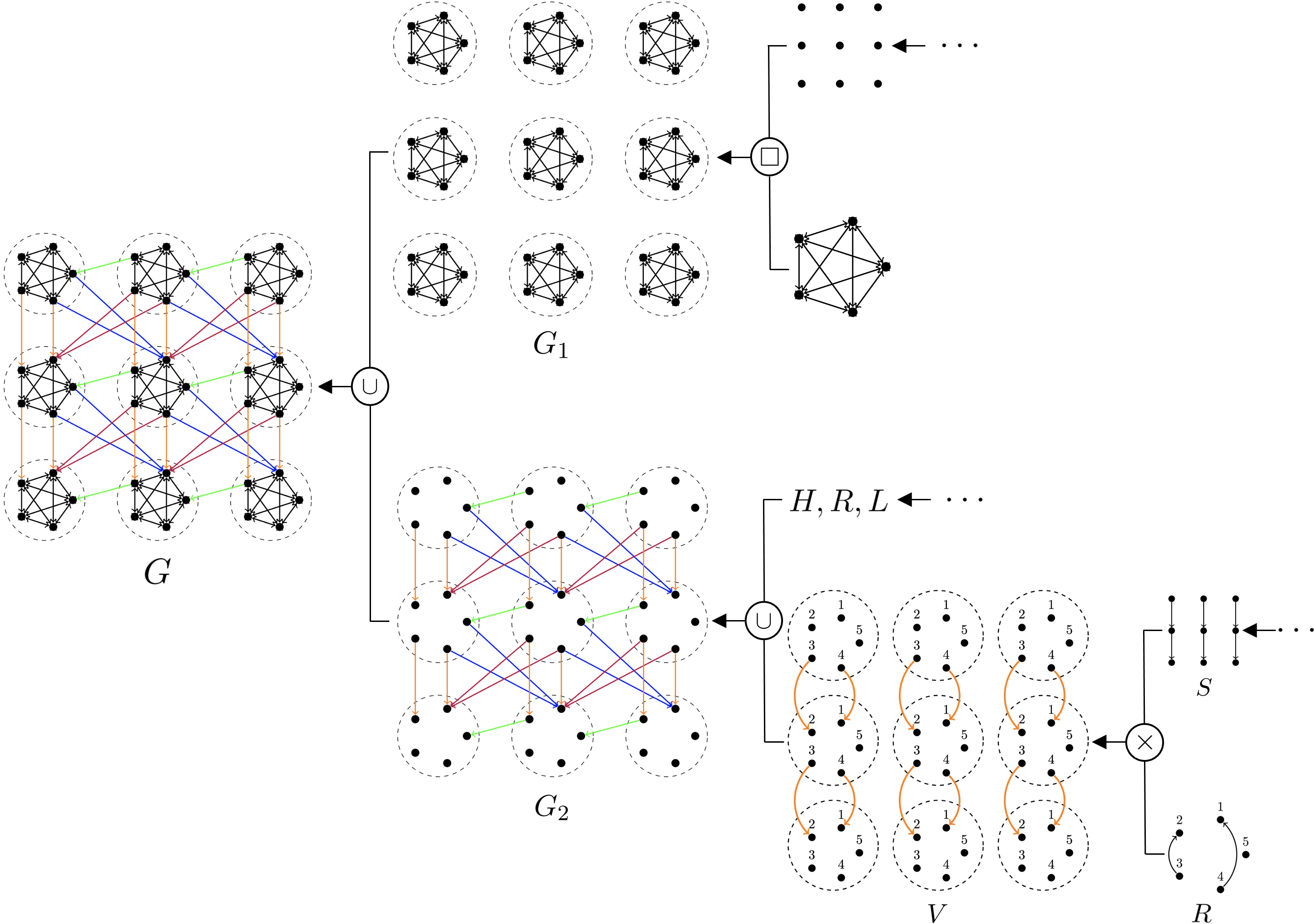}
    \end{center}
\caption{Overview of the Factorization of $G$. Supernodes are enclosed within dotted circles.}
\label{fig:LFMIS-factorization-roadmap}
\end{figure}

\paragraph*{Factored Graph Construction}
Naively, this graph has \( \Omega(T^2) \) vertices, where $T^2 = \Omega(n^{2\ell})$.
However, since the computation rules of \(M\) are local (depending only on the symbol of the work tape under the head) and repetitive (the same rules apply regardless of the head's absolute position), we can give a succinct factored representation of $G$.
Specifically, we show that \(G\) has a factored graph representation of complexity \((O(n), O(\ell^2))\).
The factorization of $G$ is outlined in \Cref{fig:LFMIS-factorization-roadmap}.

The key idea is to break down the graph $G$ into regular and repetitive substructures. 
Following this, we begin by decomposing $G$ into two subgraphs, $G_1$ and $G_2$, such that both subgraphs share the same vertex set as $G$, but $G_1$ contains only the intra-supernode edges and $G_2$ contains only the inter-supernode edges. 
$G_1$ forms a $T \times T$ grid of complete supernodes, which can be represented as the Cartesian product of an empty $T \times T$ grid of vertices and a single complete supernode. 
Similarly, \(G_2\) is a $T \times T$ grid of supernodes, but with edges connecting neighboring supernodes in four possible directions: vertical $(V)$, horizontal $(H)$, diagonally-right $R$, and diagonally-left $(L)$.
We can further decompose $G_2$ into four subgraphs $V$, $H$, $R$, and $L$, which only contain the edges which are in the direction indicated by the name of the subgraph.
Each of these subgraphs can be further decomposed into a ``structure'' graph and a ``relation'' graph using the tensor product.
The structure graph $S$ consists of a \( T \times T \) grid of (ordinary) vertices with edges connecting all neighboring vertices in the corresponding direction of the subgraph. 
The relation graph $R$ encodes the connections between vertices in neighboring supernodes in the corresponding direction. 
Moreover, the structure graphs themselves can be further factorized.
For example, the structure graph $S$ for the subgraph $V$ can be expressed as the Cartesian product of a path of length \( T \) and an empty graph on \( T \) vertices. When the path length is a perfect power \( b^k \) (where \( b, k \geq 1 \) are integers), it can be further decomposed into a union of $k$ factored graphs of complexities $(b, k)$. More details on factorization can be found in \Cref{sec:factored-graph-contruction-of-G}.

\subsubsection{Counting \texorpdfstring{$s$}{s}-Cliques}

For this result, we assume that all graphs are undirected.
In the technical overview, we use edge counting ($s = 2$) as an illustrative example.
Note that if $(u, v)$ is an edge, $u, v$ must have the same dimension (\Cref{cor:vertex-type-edge}), so we may count the edges in each dimension separately.
Thus, fix a dimension $d \leq k$.

There are at most $2^{k}$ factored components of dimension $d$.
While some edge may belong to multiple factored components, we can use the inclusion-exclusion principle to avoid double-counting such edges.
In particular, it suffices to consider counting the number of edges in an arbitrary intersection of factored components $G_{F_{1}} \cap \dotsc \cap G_{F_{m}}$.
Note that the number of such intersections, while being a large function of $k$, is crucially independent of $n$.

Fix such an intersection and consider a pair of vertices $u, v$ with $u = (u_{1}, \dotsc, u_{d})$ and $v = (v_{1}, \dotsc, v_{d})$.
Note that by determining whether $u_{i}, v_{i}$ are equal or adjacent for all $i \in [k]$, we can infer from the factored graph tree structure whether $(u, v) \in E(G)$.
Thus, we can categorize all pairs of vertices into $2^{2k}$ groups based on the relations $u_{i} = v_{i}$ and $(u_{i}, v_{i}) \in E(G_{i})$.
Note also that these groups are disjoint.
Next, we collect the subset of groups that form edges in the current intersection $G_{F_{1}} \cap \dotsc \cap G_{F_{m}}$.
To count the number of edges, it then suffices to sum up the sizes of each group in this collection.
For a given group, we can determine its size in $O(k n^{2})$ time since the relations on each coordinate are independent.
In particular, we can count the number of pairs of vertices in each input graph satisfying the relevant constraints and take the product over all input graphs.

\subsubsection{Reachability}

We begin with an overview of the proof for the $\xnl$-completeness result, followed by that for the equivalence result. 
In fact, we will see that both proofs rely on the same major building blocks in slightly different ways.

\paragraph*{$\xnl$-completeness}

We follow the standard framework for showing that reachability is $\nl$-complete \cite{sipser1996introduction}.

Membership in $\xnl$ can be established using the same algorithm that shows the ordinary reachability problem is in $\nl$, even if the input is now given as a factored graph.
As in the standard proof, we non-deterministically guess the next vertex in the path and thus only require $O(k \log n)$-space for factored graphs $G = f(G_1, \dotsc, G_k)$ of complexity $(n, k)$.
Formally, each vertex in $G$ can be specified by $O(k \log n)$ bits, since each vertex is a tuple of at most $k$ coordinates and each coordinate is a vertex in an input graph $G_i$ with at most $n$ vertices.
Thus, the algorithm requires $O(k \log n)$ bits to write down the current vertex and, since the factored graph has at most $n^{k}$ vertices, at most $O(k \log n)$ bits to keep track of the number of steps taken so far.

For $\xnl$-hardness, suppose we have some language $L \in \xnl$.
Then, there exists a nondeterministic Turing machine $M$ deciding $L$ using $f(k) \log n$-space on inputs of length $n$ and parameter $k$, for some function $f$ of $k$.
The standard reduction creates an explicit configuration graph where each vertex encodes a tuple consisting of state, input and work tape positions, and a setting of the work tape.
Since the work tape has length $f(k) \log n$, there are at least $n^{f(k)}$ distinct settings of the work tape.
Thus, even though reachability is computable in linear time, the size of the graph already depends exponentially on $f(k)$.
However, we are able to exploit the locality of Turing machine operations so that the configuration graph can in fact be encoded by a factored graph of complexity $(\poly(n), \poly(f(k)))$.
Given some configuration of $M$ on input $x$ (a tuple of state, tape head locations and work tape contents), we split the configuration into segments, where each segment only contains $\log n$ contiguous bits of the work tape contents.
Note that each segment, even if it encodes a state and tape head locations, only has $n^{O(1)}$ possible values, since one segment of the work tape only has $\log n$ bits.
Thus, we can represent all possible configurations of a segment explicitly using a graph with $n^{O(1)}$ vertices, while a product of $f(k)$ segment graphs can explicitly represent any full configuration of $M$ on input $x$.

It remains to express the appropriate transitions between configurations using a factored graph.
There are two types of transitions: \emph{intra}-segment transitions, where the work tape head stays within the same segment, and \emph{inter}-segment transitions, where the work tape head moves from one segment to another (adjacent) segment.
In either case, at most two segments of the work tape are active, where a segment is active during a transition if the work tape head either starts or ends in the segment and inactive otherwise.
Thus, we can express individual segments (or pairs of segments) explicitly using input graphs of size $n^{O(1)}$.
Since the active segments and inactive segments do not interact, we can encode all \emph{intra}-segment transitions of a single segment using factored graphs of complexity $(n^{O(1)}, f(k))$.
Similarly, we can encode all \emph{inter}-segment transitions between a single pair of adjacent segments.
Summing over all $O(f(k))$ segments and pairs of adjacent segments, we encode the configuration graph in a factored graph of complexity $(n^{O(1)}, O(f(k)^2))$.

We now briefly describe how the locality of Turing machine computation allows us to express the configuration graph of $M$ on $x$ using factored graphs.
In the overview, we describe only \emph{intra}-segment transitions.
First, note that if the work tape head is not currently placed in a segment, the work tape contents of this segment cannot change.
We thus define an \emph{inactive} graph, where for every possible work tape content, we create a self-loop vertex.
If the work tape head is currently placed in the segment, the work tape contents may change.
Thus, we define an \emph{active} graph with all possible configurations on a given segment as nodes, and edges encoding valid transitions.
For a given segment, taking the tensor product of the active graph for this segment and the inactive graph for all other segments encodes all \emph{intra}-segment transitions.
A similar construction can be used to construct factored graphs that encode \emph{inter}-segment transitions.

\paragraph*{Equivalence to the Open Problem}

For the forward direction, suppose there exists an absolute constant $C$ such that $\nl \subseteq \dtime(n^C)$.
To show that $\xnl \subseteq \fpt$, it now suffices to show reachability on factored graphs is in $\fpt$ due to its $\xnl$-completeness under $\fpt$-reductions.
Note that the algorithm used in the proof of $\xnl$-membership is in fact an $\nl$ algorithm for each fixed parameter $k$.
Thus, if $\nl \subseteq \dtime \left( n^{C} \right)$ for some $C$ independent of $k$, then for any fixed $k$, reachability on factored graphs of complexity $(n, k)$ can be computed in time $O(k^{C} n^{2 C})$ (as factored graphs of this complexity have input size $O(k n^2)$) and is therefore in $\fpt$.

Conversely, if $\xnl \subseteq \fpt$, then in particular, reachability on factored graphs is in $\fpt$.
Now, consider any language $L \in \nl$ with its associated nondeterministic Turing machine $M$ that decides $L$ in $S \log n$ space for some constant $S$. Using the same construction as in the proof of $\xnl$-hardness, we construct a factored graph $G$ of complexity $(n^{O(1)}, O(S^2))$ where solving reachability on $G$ simulates $M$.
Since reachability on factored graphs is in $\fpt$, it follows that reachability on $G$ can be solved in time $O(f(S^2) n^{O(1)})$ for some function $f$, where the exponent on $n$ is independent of $S$, hence independent of the specific language $L \in \nl$.
This shows that $\nl \subseteq \dtime(n^C)$ for some absolute constant $C$.

\subsection{Related Work}

A long line of work, initiated by \cite{downey1995fixed, downey1995fixed2, downey1996descriptive, flum2001fixed}, has studied the complexity of parameterized problems as a function of their input size and a parameter.
Within parameterized complexity, a common theme is to study the complexity of problems given succinct representations of their input.
For example, several previous works have investigated the complexity of computing Nash equilibrium of \emph{succinct} games (represented implicitly) \cite{feigenbaum1995game, daskalakis2005games, daskalakis2006gameflat, fortnow2008zerosum, papadimitriou2008computing, greco2015compact}.
As another example, \cite{wagner1986complexity} considers the complexity of combinatorial problems with succinct representation.
Similarly, \cite{orlin1981complexity, wanke1993paths, hofting1993polynomial, hoppe2000quickest, marathe1994approximation, marathe1998theory, chen2003periodic, chen2005periodic} study the complexity of various computational problems on periodic structures, i.e., travel schedules on a periodic timetable.

Most relevant to this paper, several works have investigated computational problems on graphs with succinct representations such as small circuits \cite{galperin1983succinct}, distributed graphs described by low complexity agents \cite{arora2009lowcomplexitygraphs}, and factored problems \cite{dalirrooyfard2020factor}.
However, none of these works consider the complexity of factored graphs formed under graph products.

On the subject of succinct representations, researchers have also studied how to represent graphs as efficiently as possible \cite{kannan1988implicit, he1999planar, barbay2007succinct, farzan2008succinct, blelloch2010succinct, maneth2015survey, parque2017succinct}.
% \cite{viola2018localexpanders} constructs (expander) graphs whose neighborhoods can be computed by maps with constant locality. 

\subsection{Outline}

In \Cref{sec:prelims}, we provide formal definitions of the relevant graph operations and the construction of factored graphs.
We also review key concepts in parameterized complexity that are relevant to our work.
In \Cref{sec:LFMIS}, we show that LFMIS is $\xp$-complete and therefore unconditionally not in $\fpt$.
In \Cref{sec:counting-subgraphs}, we show that counting small cliques is in $\fpt$.
In \Cref{sec:reachability}, we show that reachability is $\xnl$-complete and present a condition under which $\xnl$ is (or is not) contained in $\fpt$.

\section{Preliminaries}
\label{sec:prelims}

Unless otherwise noted, we work with directed graphs, with edges from $a$ to $b$ denoted by $(a, b)$.
For a graph $G$, we denote its vertices by $V(G)$ and edges by $E(G)$. 
For any subset of vertices $S \subset V(G)$, let $G[S]$ denote the subgraph induced by $S$.
For a set $X$, we define $\cP(X) = \{Y : Y \subseteq X\}$ to be the power set of $X$.

\subsection{Graph Operations}
\label{sec:graph-products}
We begin with the definitions of the relevant graph operations.
Let $G$ and $H$ be two directed graphs. 

\begin{definition}
    \label{def:cartesian-prouct}
    The \emph{Cartesian product} \(G \cart H\) of \(G\) and \(H\) has vertex set  \(V(G) \times V(H)\) and directed edges \(((v_1, u_1), (v_2, u_2))\) if and only if either
    \begin{itemize}
        \item \(v_1 = v_2\) and \((u_1, u_2) \in E(H)\), or
        \item \(u_1 = u_2\) and \((v_1, v_2) \in E(G)\)
    \end{itemize}
\end{definition}

As a simple example, note that the Cartesian product of two paths is a grid.

\begin{definition}
    \label{def:tensor-product}
    The \emph{tensor product} \(G \times H\) of $G$ and $H$ has vertex set \(V(G) \times V(H)\) and directed edges \(((v_1, u_1), (v_2, u_2))\) if and only if \((v_1, v_2) \in E(G)\) and \((u_1, u_2) \in E(H)\).
\end{definition}

\begin{definition}
    \label{def:graph-union}
    The \emph{union} \(G \cup H\) of $G$ and $H$ has vertex set $V(G) \cup V(H)$ and edge set $E(G) \cup E(H)$.
\end{definition}

While there are many other graph products and operations to consider, such as the or product, or graph negation, we will primarily focus our study of factored graphs under the above three operations.

We note that the above three operations are associative, and observe that the products of higher arity are given as follows.

\begin{definition}
    \label{def:cartesian-prouct-k}
    The \emph{Cartesian product} \(\cart_{i = 1}^{k} G_{i}\) has vertex set \(\prod_{i = 1}^{k} V(G_{i})\) and directed edges \(((v_1, \dotsc, v_{k}), (u_1, \dotsc, u_{k}))\) if and only if there is some index $j \in [k]$ such that $(v_{j}, u_{j}) \in E(G_{j})$ and $v_{i} = u_{i}$ for all $i \neq j$.
\end{definition}

\begin{definition}
    \label{def:tensor-product-k}
    The \emph{tensor product} \(\times_{i = 1}^{k} G_{i}\) has vertex set \(\prod_{i = 1}^{k} V(G_{i})\) and directed edges \(((v_1, \dotsc, v_{k}), (u_1, \dotsc, u_{k}))\) if and only if \((v_i, u_i) \in E(G_{i})\) for all $i \in [k]$.
\end{definition}

\begin{definition}
    \label{def:graph-union-k}
    The \emph{union} \(\bigcup_{i = 1}^{k} G_{i}\) of $G$ and $H$ has vertex set \(\bigcup_{i = 1}^{k} V(G_{i})\) and edge set \(\bigcup_{i = 1}^{k} E(G_{i})\).
\end{definition}

\subsection{Factored Graph Construction}
\label{sec:factored-graph-prelims}

We now describe how the above operations construct factored graphs.
Formally, we define a factored graph by describing how the above graph operations combine the input graphs into a single graph.
This motivates the definition of a factored graph according to a tree which specifies the order of operations, or the tree structure of a factored graph.

\begin{definition}[Factored Graph Tree Structure]
    \label{def:factored-graph-tree}
    The \emph{factored graph tree structure} with $k$ leaves, denoted $f(G_{1}, \dotsc, G_{k})$, is an ordered tree with $k$ leaves where internal nodes are labelled by an operation (one of $\square, \times$, or $\cup$) and the leaves are labelled by graphs $G_{1}, \dotsc, G_{k}$.
    Note that $G_{i}$ are arbitrary and not necessarily distinct.
    We require that each internal node has degree at least $2$.

    Let $G = f(G_1, \dotsc, G_{k})$ denote that $G$ is the \emph{factored graph} given by the factored graph tree structure $f(G_1, \dotsc, G_{k})$.
    For convenience, let $T(G)$ denote the factored graph tree structure.
    We say that the factored graph tree structure $f(G_1, \dotsc, G_{k})$, or simply $G$, is \emph{of complexity $(n, k)$} if $T(G)$ has at most $k$ leaves and each leaf is labelled by a graph with at most $n$ vertices.
\end{definition}

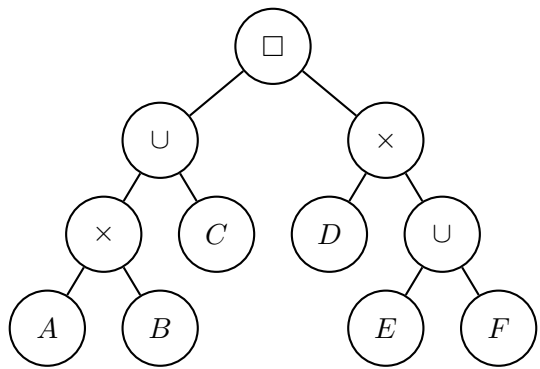
\begin{figure}[ht]
    \centering
    
    \tikzstyle{node}=[circle, draw, minimum size=10mm, inner sep=0pt]
    \tikzstyle{root}=[ellipse, draw, minimum width=2cm, minimum height=1cm, inner sep=0pt, align=center, text width=2cm]

    \begin{tikzpicture}[thick, level distance=1.25cm, level 1/.style={sibling distance=3cm}, level 2/.style={sibling distance=1.5cm}]
        \node[node] {$\square$}
            child {node[node] {$\cup$}
                child {node[node] {$\times$}
                    child {node[node] {$A$}}
                    child {node[node] {$B$}}
                }
                child {node[node] {$C$}}
            }
            child {node[node] {$\times$}
                child {node[node] {$D$}}
                child {node[node] {$\cup$}
                    child {node[node] {$E$}}
                    child {node[node] {$F$}}
                }
            };
    \end{tikzpicture}
    
    \caption{Tree structure of the factored graph $G = ((A \times B) \cup C ) \cart (D \times (E \cup F))$.}
    \label{fig:vertex-type-tree-structure}
\end{figure}

To illustrate our definitions, we will use the following example of a factored graph \( G = ((A \times B) \cup C) \cart (D \times (E \cup F)) \).
Note that there is a natural correspondence between the tree structures of factored graphs and the formulas of the above form, by using parentheses to delineate subtrees of the tree structure.
The tree structure of $G$ is given in \Cref{fig:vertex-type-tree-structure}.

We now describe how the vertex and edge sets of a factored graph are obtained from its tree structure.
First, we define factored components of factored graphs.

\begin{definition}[Factored Component]
    \label{def:factored-components}
    Let $G$ be a factored graph.
    A \emph{factored component} $G_{F}$ of $G$ is the factored graph whose tree structure $T(G_{F})$ is obtained from the tree structure of $G$ by recursively replacing each internal node labelled by $\cup$ with the subtree rooted at one of its children.
    Note that the internal nodes of $T(G_{F})$ are labelled either $\cart$ or $\times$.

    We say $G_{F}$ has \emph{dimension} $\ell$ if $T(G_{F})$ has $\ell$ leaves.
    Suppose the leaves are labelled $G_{1}, \dotsc, G_{\ell}$.
    Then, $V(G_{F}) = \prod_{i = 1}^{\ell} V(G_{i})$.
    We say every vertex $v \in V(G_{F})$ has dimension $\ell$.
    The edge set $E(G_{F})$ is determined by $T(G_{F})$ and $E(G_1), \dotsc, E(G_{\ell})$ following the rules of Cartesian and tensor products.
\end{definition}

In our example, the factored components are $(A \times B) \cart (D \times E), (A \times B) \cart (D \times F), C \cart (D \times E)$, and $C \cart (D \times F)$, with dimensions $4, 4, 3, 3$, respectively.
Some example tree structures of factored components are given in \Cref{fig:vertex-tree-types}.

\begin{figure}[ht]
    \centering

    \tikzstyle{node}=[circle, draw, minimum size=10mm, inner sep=0pt]
    \usetikzlibrary{shapes.geometric}

    \begin{minipage}[t]{0.45\textwidth}
        \centering
        \begin{tikzpicture}[thick, level distance=1.25cm, level 1/.style={sibling distance=3cm}, level 2/.style={sibling distance=1.5cm}]
        \node[node] {$\square$}
            child {node[node] {$\times$}
                child {node[node] {$A$}}
                child {node[node] {$B$}}
            }
            child {node[node] {$\times$}
                child {node[node] {$D$}}
                child {node[node] {$E$}}
            };
    \end{tikzpicture}
        \caption*{Tree structure of $(A \times B) \cart (D \times E)$.}

    \end{minipage}
    \hfill
    \begin{minipage}[t]{0.45\textwidth}
        \centering
        \begin{tikzpicture}[thick, level distance=1.25cm, level 1/.style={sibling distance=3cm}, level 2/.style={sibling distance=1.5cm}]
        \node[node] {$\square$}
            child {node[node] {$C$}}
            child {node[node] {$\times$}
                child {node[node] {$D$}}
                child {node[node] {$F$}}
            };
        \end{tikzpicture}
        \caption*{Tree structure of $C \cart (D \times F)$.}
        
    \end{minipage}

\caption{Examples of the Tree Structures of Factored Components.}
\label{fig:vertex-tree-types}
\end{figure}

Here, we emphasize that the vertex sets of the factored components are \emph{flattened} products of the vertex sets of the graphs labeling the leaves.
For example, the vertex of the factored component $(A \times B) \cart (D \times E)$ has the form $(a, b, d, e)$ not $((a, b), (d, e))$.
In particular, the vertex sets do not depend on the topology of the tree beyond the order of the leaves.
Finally, the vertex and edge set of $G$ is simply the union of the factored components.

\begin{definition}[Factored Graph]
    \label{def:factored-graph-edge-set}
    Let $G = f(G_{1}, \dotsc, G_{k})$ be a factored graph with tree structure $T(G)$ and factored components $G_{F_{1}} , \dotsc, G_{F_{m}}$.
    Define $V(G) = \bigcup_{i = 1}^{m} V(G_{F_{i}})$ and $E(G) = \bigcup_{i = 1}^{m} E(G_{F_{i}})$.
\end{definition}

We note here that different factored graphs can have identical vertex and edge sets.
For example, $G_1 = (H_1 \times H_2) \times H_3$ and $G_2 = H_1 \times (H_2 \times H_3)$ both have vertex set $\prod_{i = 1}^{3} H_{i}$ and edges from $(h_{1}, h_{2}, h_{3})$ to $(h_{1}', h_{2}', h_{3}')$ if and only if $(h_{i}, h_{i}') \in E(H_{i})$ for all $i$.
This aligns with our expectation, as the tensor product is associative.

Note that a single vertex in $G$ can belong to multiple factored components.
In our running example, if there is a vertex $x \in V(E) \cap V(F)$, then the vertex $(c, d, x)$ is in the factored components corresponding to $C \cart (D \times E)$ and $C \cart (D \times F)$.
However, both factored components have dimension $3$.
This is formalized in the following lemma, which follows directly from the construction of factored graphs.

\begin{lemma}
    \label{lemma:factored-component-same-dimension}
    Suppose $v \in V(G)$ is in factored components $G_{F}, G_{F'}$.
    Then, $\dim(G_{F}) = \dim(G_{F'})$.
\end{lemma}

This allows us to define the dimension of a vertex $\dim(v)$ as the dimension of any factored component it belongs to.

\begin{definition}[Vertex Dimension]
    \label{def:vertex-dimension}
    Let $G = f(G_1, \dotsc, G_m)$ be a factored graph and $v \in V(G)$ be a vertex.
    The \emph{dimension} of $v$, denoted by $\dim(v)$, is the dimension $\dim(G_F)$ of any factored component $G_{F}$ that contains $v$.
\end{definition}

By construction, the endpoints of any edge must belong to the same factored component.

\begin{lemma}
    \label{lemma:factored-component-edges}
    Let $(u, v) \in E(G)$ be an edge.
    Then, $u$ and $v$ belong to the same factored component.
\end{lemma}

As a corollary, we show that there are no edges between vertices of different dimensions, since vertices in the same factored component necessarily have the same dimension.

\begin{corollary}
    \label{cor:vertex-type-edge}
    Let $(u, v)$ be an edge in $G$.
    Then, $u$ and $v$ have the same dimension.
\end{corollary}

Finally, since every vertex has a unique dimension, it is clear that vertex dimension partitions the vertices of the factored graph.
Thus, vertex dimension induces an equivalence relation on $V(G)$.

\subsection{Parameterized Complexity}\label{sec:param-complexity}
In this section, we review the parameterized complexity classes relevant to this paper with respect to problems on factored graphs.
We begin with the definition of fixed-parameter tractability ($\fpt$).

\begin{definition}[Fixed-parameter Tractability ($\fpt$) for Factored Graph Problems]
    \label{def:fpt}
    A problem on factored graphs is \emph{fixed-parameter tractable} if there exists an algorithm that solves the problem on factored graph inputs of complexity $(n, k)$ in time $O\left(g(k) n^{C}\right)$, where $g$ is a function of $k$ and $C$ is a constant independent of $k$.
\end{definition}

We also study the parameterized complexity classes $\xp$ and $\xnl$, which correspond to the parameterized version of the classical complexity classes $\ptime$ and $\nl$, respectively.
More generally, one can define the parameterized version $\mathbf{XC}$ for any classical complexity class $\mathbf{C}$ as follows \cite{downey2012parameterized}:
a parameterized language $L \in \mathbf{XC}$ if and only if $L_k \in \mathbf{C}$ for every parameter $k$.
Here, $L_k$ is referred to as the \emph{$k$-th slice} of $L$, which is the subset of $L$ consisting of all instances with parameter $k$.
We now give the definitions of $\xp$ and $\xnl$ in the context of factored graph problems.

\begin{definition}[$\xp$ for Factored Graph Problems]\label{def:xp}
    A problem on factored graphs is \emph{in $\xp$} if there exists an algorithm that solves the problem on factored graph inputs of complexity $(n, k)$ in time $O\left(f(k) n^{f(k)}\right)$, where $f$ is a function of $k$.
\end{definition}

\begin{definition}[$\xnl$ for Factored Graph Problems]\label{def:xnl}
    A problem on factored graphs is \emph{in $\xnl$} if there exists a nondeterministic algorithm that solves the problem on factored graph inputs of complexity $(n, k)$ in space $O\left(f(k) \log n\right)$, where $f$ is a function of $k$.
\end{definition}

Finally, for $\xp$- and $\xnl$-completeness, we follow the usual definition that a problem is considered to be \emph{complete} for a complexity class \textbf{C} if it belongs to \textbf{C} and is \emph{hard} for \textbf{C} under some suitable notion of reductions.
Here, we use the standard notion of parameterized mapping reduction (see Definition 9.3 in \cite{downey2012parameterized}), also known as the $\fpt$-reduction \cite{chen2003bounded-nondeterminism-param-complexity}, under which the class $\fpt$ is closed.
We now give a definition for $\xp$- and $\xnl$-hardness under this reduction in the context of factored graph problems.

\begin{definition}[$\xp$-hardness (resp. $\xnl$-hardness) for Factored Graph Problems]\label{def:xp-xnl-complete}
    A problem $L'$ on factored graphs is \emph{$\xp$-hard} (resp. \emph{$\xnl$-hard}) if for every parameterized language $L \in \xp$ (resp. $L \in \xnl$), there exists a mapping $F$ such that for every parameter $k$, $L_k$ mapping reduces to $L'_{k'}$ under $F$ using $O(f(k) n^{O(1)})$ time, where $f$ is a function of $k$ and $k'$ depends only on $k$.
\end{definition}

\section{Lexicographically First Maximal Independent Set on Factored Graphs}
\label{sec:LFMIS}

In this section, we show that the decision version of the lexicographically first maximal independent set problem (LFMIS) defined on factored graphs is $\xp$-complete.
In particular, this means that it is \emph{unconditionally} not in $\fpt$, and we show an explicit lower bound of $n^{\Omega(\sqrt{k})}$ on factored graphs of complexity $(n, k)$.
We begin with a formal definition of the factored version of the LFMIS problem.

\subsection{Problem Definition}

In this section, we formalize the parameterized version of the LFMIS problem on factored graph inputs.
Let us first recall the basic definitions involved in the standard LFMIS problem on ordinary graphs.

\begin{definition}
    Let $G = (V(G), E(G))$ be a directed graph. A subset of vertices $I \subseteq V$ is called an \emph{independent set} if for every pair of vertices \( u, v \in I \), we have $(u, v) \notin E(G)$ and $(v, u) \notin E(G)$. An independent set \( I \) is called \emph{maximal} if whenever $I \subsetneq J \subseteq V$, $J$ is not an independent set.

    If $V(G) = \{0, 1, \dotsc, |V(G)| - 1\}$, then for each maximal independent set $I$ of $G$, we associate the sequence formed by placing all the vertices of $I$ in increasing order. We say that the maximal independent set $I$ is \emph{lexicographically first} if its associated sequence has the lowest lexicographic order among all sequences corresponding to maximal independent sets. In this case, we denote $I = \lfmis(G)$.
\end{definition}

In this paper, we focus on the decision version of the lexicographically first maximal independent set problem.

\begin{definition}[Lexicographically First Maximal Independent Set Problem (LFMIS)]
     Given a directed graph $G = (V(G), E(G))$ with $V(G) = \{0, 1, \dotsc , |V(G)| - 1\}$, and a vertex $s \in V(G)$, decide whether $s \in \lfmis(G)$. 
\end{definition}

Here, we briefly review the greedy algorithm, which correctly returns the LFMIS of an input graph:

\begin{algorithm}
\caption{Greedy Algorithm for Computing LFMIS}\label{alg:greedy}
\begin{algorithmic}
\Require $G = (V(G), E(G))$ with $V(G) = \{0, 1, \dotsc, |V(G)| - 1\}$
\Ensure $\lfmis(G)$
\State $I \gets \emptyset$
\For{$v \gets 0$ to $|V(G)| - 1$}
    \If{$v$ does not form an edge with any $w \in I$}
        \State $I \gets I \cup \{v\}$
    \EndIf
\EndFor
\State \Return $I$
\end{algorithmic}
\end{algorithm}

We may think of $V(G) = \{0, 1, \dotsc, |V(G)| - 1\}$ as defining an ordering on the vertex set. Then in words, the greedy algorithm iterates through the vertex set $V(G)$ according to this ordering from $0$ to $|V(G)| - 1$, and adds a vertex to the independent set as long as it does not form an edge with any previously added vertices. Therefore, we think of a vertex $u$ as having \emph{higher priority} than $v$ if $u$ has a lower index than $v$. From the greedy algorithm, we observe the following characterization of the LFMIS of a graph:

\begin{proposition}[Characterization of LFMIS]\label{prop:characterization-of-LFMIS}
    Let $G = (V(G), E(G))$ be a directed graph with $V(G) = \{0, 1, \dotsc, |V(G)| - 1\}$. A vertex $v \in V(G)$ belongs to $\lfmis(G)$ if and only if for all vertices $u \in \lfmis(G)$ with a higher priority than $v$, there is no edge between $u$ and $v$. 
\end{proposition}

Now, we define the version of the LFMIS problem on factored graphs.
The LFMIS problem on factored graphs is exactly the same as the standard version, except that the input is a factored graph $G = f(G_1, \dotsc, G_k)$, with indices $V(G_i) = \set{0, 1, \dotsc, V(G_i) - 1}$ given only for each $G_i$. 
Unlike the standard LFMIS problem, which explicitly provides the indices for the vertex set $V(G)$ as input, we need to define a way in which the factored version considers the indices on $V(G)$ given only the indices for each $V(G_i)$.
Recall that each vertex in the factored graph $G$ is a flattened tuple of numbers $(v_1, \dotsc, v_{k'})$ for some $k' \leq k$, and therefore, we define the vertex indices on $V(G)$ to be given according to the standard lexicographic order of these tuples, with the index 0 given to the vertex with the lowest lexicographic order.

\subsection{LFMIS and Computation in Polynomial Time}

Recall that our first main result establishes the $\xp$-completeness of the LFMIS problem on factored graphs under $\fpt$-reductions:

    \LFMISXPC*

\noindent It turns out that \Cref{thm:factored-LFMIS-XP-complete} is a consequence of the following key reduction lemma:

\begin{lemma}\label{lem:factored-LFMIS-reduction}
    Let $L$ be a language in $\dtime(n^{\ell})$.
    Then, $L$ is mapping reducible using $O(\ell^2 n^2)$ time to the LFMIS problem on factored graphs with parameter $O(\ell^2)$.
    In particular, the reduction maps every input of length $n$ to a factored graph of complexity $(O(n), O(\ell^2))$.
\end{lemma}

The goal of this reduction is to construct a grid-like graph $G$ that encodes the computation of a Turing machine.
However, an explicit representation of $G$ could require up to $\Omega(n^{2\ell})$ space in order to span the computational time and space of an $O(n^\ell)$-time Turing machine.
The key to proving \Cref{lem:factored-LFMIS-reduction} is to leverage the locality of Turing machine computations and identify repetitive substructures of $G$, which allows us to represent $G$ in a compressed fashion using the graph operations in \Cref{sec:graph-products}.
Now, let us first demonstrate how \Cref{thm:factored-LFMIS-XP-complete} directly follows from \Cref{lem:factored-LFMIS-reduction}.

\begin{proof}[Proof of \Cref{thm:factored-LFMIS-XP-complete}]
        We first show that the LFMIS problem on factored graphs is $\xp$-complete.
        Given an input factored graph of complexity $(n, k)$, one can compute the explicit representation of the factored graph, which takes $O(n^k)$ time, and then apply the greedy algorithm in polynomial time.
        Thus, the LFMIS problem on factored graphs is in $\xp$.
        To show $\xp$-hardness, consider an arbitrary language $L$ in $\xp$. There exists a function $f$ such that the language $L_k \in \dtime(n^{f(k)})$ for each parameter $k \in \N$.
        Then, \Cref{lem:factored-LFMIS-reduction} proves the $\xp$-hardness by choosing $\ell = f(k)$ according to \Cref{def:xp-xnl-complete}, and we conclude that the LFMIS problem on factored graphs is $\xp$-complete.
        
        For the lower bound, suppose for the sake of contradiction that the LFMIS problem on an input factored graph $G$ of complexity $(n, k)$ can be solved in $n^{o(\sqrt{k})}$ time. 
        Due to the Time Hierarchy Theorem, there exists a language $L$ that can be decided in $O(n^k)$ times but cannot be decided in $O(n^{k - \varepsilon})$ times for any $\varepsilon > 0$. 
        Let $M$ be the Turing machine that decides $L$ running in time $O(n^k)$. 
        By \Cref{lem:factored-LFMIS-reduction}, we can simulate $M$ on input of size $n$ by solving the LFMIS problem on the input factored graph of complexity $(O(n), O(k^2))$. But this means the simulation can be done within $n^{o(k)}$ time, which is a contradiction. 
    \end{proof}

The remainder of \Cref{sec:LFMIS} is devoted to the proof of the key reduction \Cref{lem:factored-LFMIS-reduction}.
In \Cref{sec:LFMIS-reduction-motivation}, we begin by discussing the intuition behind the proof and introducing the key components of the reduction.
Specifically, we explain how to define a graph $G$ such that the process of solving the LFMIS on $G$ can be considered as recovering the ``computation history matrix'' of a Turing machine, thereby simulating its computation.
In \Cref{sec:LFMIS-reduction}, we give an explicit definition of the graph $G$ used in the reduction and prove its correctness.
Following this, in \Cref{sec:factored-graph-contruction-of-G}, we provide an efficient factored graph representation of $G$ with the specified complexity, thus completing the proof of the key reduction \Cref{lem:factored-LFMIS-reduction}.

\subsection{Proof Setup}\label{sec:LFMIS-reduction-motivation} 

In this section, we present the intuition behind our proof and introduce all the necessary key components for the reduction.
The reduction we use is a generic reduction that involves working with the Turing machine that decides a language.
In particular, the reduction maps an input to the Turing machine to an instance of the LFMIS problem on factored graphs such that solving the LFMIS on the factored graph instance simulates the computation of the Turing machine.
Therefore, we begin with the definition of the Turing machine model we use in the reduction with some additional assumptions.

\subsubsection{Turing Machine Model and Assumptions}\label{sec:Turing-machine-model}

In this reduction, we use the standard deterministic single-tape Turing machine model.
A Turing machine is given as \(M = \langle Q, \Gamma , \Sigma , \delta , q_0, \qacc, \qrej \rangle \), where \(Q\) is the set of states, \(\Gamma\) is the set of tape alphabets, \(\Sigma \subseteq \Gamma  \setminus  \{\perp\}\) is the set of input alphabets, $\perp \in \Gamma$ is the blank symbol, \(\delta : Q \times \Gamma  \to Q \times \Gamma  \times \{L, R\}\) is the transition function, \(q_0 \in Q\) is the start state, \(\qacc \in Q\) is the accept state, and \(\qrej \in Q\) is the reject state.
We use \(F := \{\qacc, \qrej\} \subseteq Q\) to denote the set of \emph{halting states}.

For convenience, we use the notations $\delta_Q, \delta_{\Gamma}$, and $\delta_D$ to denote the components of the transition function output.
The notation $\delta(q, a) = (\delta_Q(q, a), \delta_{\Gamma}(q, a), \delta_D(q, a))$ indicates that when the Turing machine reads the symbol $a$ in state $q$, it transitions to state $\delta_Q(q, a)$, writes $\delta_{\Gamma}(q, a)$ on the current tape cell, and moves the tape head in direction $\delta_D(q, a) \in \{L, R\}$ ($L$ indicates left and $R$ indicates right).
Moreover, we make the following assumptions about $M$.
\begin{assumption}\label{assumption:tm-left-endmarker}
    We assume that there is always a left endmarker $\$$ in the first tape cell, and the tape head always moves right whenever it reads $\$$.
\end{assumption}
\noindent This ensures that the tape head always moves in the direction indicated by the transition function. (In the standard Turing machine model, the tape head could get stuck in the same tape cell if it tries to move left while over the leftmost tape cell.) 
\begin{assumption}\label{assumption:tm-tapehead-move-to-the-front-when-terminating}
    We assume that when \(M\) decides to transition into a halting state, the tape head first moves all the way to the left endmarker, overwriting all the symbols with $\perp$ along the way, and then it moves right and stops over the second tape cell in the corresponding halting state.
\end{assumption}
\noindent We note that both assumptions above would only slow down the running time by a constant factor.

\subsubsection{Computation History Matrix of a Turing Machine}\label{sec:computational-history-matrix}

\paragraph*{Defining the Computation History Matrix}
Let $L$ be a language and $M$ be a Turing machine that decides $L$. 
Let $x = (x_1, \dotsc, x_n)$ be an input of size $n$ and $T$ be the running time of $M$ on $x$. 
We now describe how to encapsulate the entire computation history of $M$ on $x$ in a $T \times T$ matrix $W$, where the $i$-th row of $W$ represents the configuration of $M$ at time $i$.

Note that $M$ cannot use more than $T$ tape cells, since $M$ halts in $T$ steps and the head is not allowed to move for more than one cell to the left or right.
Thus, a configuration of $M$ at some time $i$ includes 1) the string $a_1 a_2 \dotsc a_T$ on the current tape, 2) the machine's current state $q$, and 3) the position $j$ of the tape head.
A typical way to represent this configuration is
\[
    a_1 a_2 \dotsc a_{j-1} (q a_j) a_{j+1} \dotsc a_T.
\]
 Now, we insert a dummy state symbol $*$ in front of the $a_{j'}$'s that are not pointed by the tape head to make the representation uniform, and chunk the representation into state-alphabet pairs
\[
    (*, a_1), (*, a_2), \dotsc, (*, a_{j-1}), (q, a_j), (*, a_{j+1}), \dotsc, (*, a_T).
\]
We define this to be the $i$-th row of the matrix $W$.
Formally, each entry of $W$ is a pair in the set $Q^* \times \Gamma$, where $Q^* := Q \cup \{*\}$.
For each $1 \leq i, j \leq T$, if $a \in \Gamma$ is the symbol occupying the $j$-th cell and $q \in Q$ is the state at time $i$, then we define
\[
    W_{i, j} = 
    \begin{cases}
        (q, a) & \text{ if the head is over the $j$-th cell at time $i$} \\
        (*, a) & \text{ otherwise}
    \end{cases}
\]
If $M$ has already halted before time $i$, we assume that the configuration of $M$ at time $i$ stays at the halting configuration.

For technical reasons, in order to simulate the Turing machine by solving the LFMIS problem on a graph, we define a special start state $\hat{q}_0$, a special endmarker $\hat{\$}$, and $n$ special input alphabets $\hat{x}_1, \hat{x}_2, \dotsc, \hat{x}_n$. The first row of $W$ is now replaced with
\[
    (\hat{q}_0, \hat{\$}), (*, \hat{x}_1), \dotsc, (*, \hat{x}_n), (*, \perp), \dotsc, (*, \perp),
\]
but all the other rows remain unchanged. Finally, we define the set of all possible entries as $\tiles$, where $\hat{Q} := Q^* \cup \{\hat{q}_0\}$ and $\hat{\Gamma} := \Gamma \cup \{\hat{\$}, \hat{x}_1, \hat{x}_2, \dotsc, \hat{x}_n\}.$

\begin{remark}
    We note that even if the input alphabets might repeat, i.e. $x_i = x_j$ for some $i \neq j$, we define their special alphabets as \emph{distinct} elements $\hat{x}_i \neq \hat{x}_j$.
\end{remark}

It turns out that there is a natural construction of a graph $G$ such that solving the LFMIS problem on $G$ exactly recovers all the entries of the computation history matrix $W$.

\paragraph*{Recovering the Computation History Matrix by Solving LFMIS}

In this reduction, our goal is to define a graph $G$ such that solving the LFMIS problem on $G$ recovers all the entries of the computation history matrix $W$, thereby simulating the Turing machine $M$. We provide the motivation behind the definition of $G$ in this section.

The graph $G$ has a $T \times T$ grid structure that mirrors the layout of the matrix $W$. 
At each grid point, we place a copy of the set of all possible matrix entries $\tiles$. (Here, we think of each element of $\tiles$ as a vertex.) 
We refer to each of these copies as a \emph{supernode} and denote the supernode at the $i$-th row and $j$-th column by $S_{i, j}.$
Intuitively, the supernode $S_{i, j}$ represents all possible choices for the matrix entry $W_{i, j}$, and the goal is to ensure that the LFMIS of $G$ contains exactly the one vertex in $S_{i, j}$ that equals $W_{i, j}.$

Since the LFMIS is an independent set, our strategy is to define the edges of $G$ in a way that guides the selection of vertices for the LFMIS. 
It is fairly straightforward to ensure that the LFMIS contains at most one vertex from each supernode, since we can simply construct a complete directed graph on each supernode. 
The challenge, however, is to ensure that the LFMIS contains the correct vertex from each supernode that matches the corresponding entry in $W$.
To achieve this, we leverage the deterministic behavior of the Turing machine: the configuration at any point during the computation uniquely determines the next configuration according to the transition function. 
Therefore, we define downward-directed edges between supernodes in adjacent rows.
These edges represent \emph{inconsistencies}: specifically, for $v \in S_{i, j}$ and $v' \in S_{i + 1, j'}$, a directed edge $(v, v')$ indicates that, if $v$ is chosen to represent the $j$-th tape cell at time $i$, then $v'$ cannot represent the $j'$-th tape cell at the next time step $i + 1$, based on the machine's transition function.
Using the characterization from \Cref{prop:characterization-of-LFMIS}, we can make the LFMIS consider the supernodes in a row-by-row order by defining a proper index on the vertex set.
Thus, if we can ensure that the LFMIS correctly recovers the first row of the computation history $W$, then it can inductively recover the entire computation history by choosing vertices from subsequent rows that are consistent with the choices in the previous row.

Now, recall that the tape head moves at most one step left or right at a time.
As a result, determining each one entry does not require the knowledge of the entire previous configuration.
Instead, it suffices to define edges only from the (up to) three \emph{neighboring} supernodes in the previous row. 
This leads to the definition of \emph{local consistencies}.

\subsubsection{Local Consistencies of Turing Machine Computation}\label{sec:local-consistencies}
In this section, we define the \emph{consistency functions} that are necessary for the LFMIS to contain the correct vertex from each supernode.
A pair of supernodes in the grid can be neighboring in one of four directions: vertical $(V)$, horizontal $(H)$, diagonally-right $(R)$, and diagonally-left $(L)$. We use $\cD := \{V, H, R, L\}$ to denote the set of possible neighbor directions.
For each supernode, we consider the following neighbors, which we call \emph{parents}:

\begin{definition}\label{def:parent}
    Let $(i, j), (i', j') \in \Z^2$ be two integer coordinates. We say that $(i, j)$ is the \emph{parent} of $(i', j')$ in direction $D$ if 
    \[  
    (i', j') = 
        \begin{cases}
           (i+1, j) &\text{ if } D = V \\
           (i, j+1) &\text{ if } D = H \\
           (i+1, j+1) &\text{ if } D = R \\
           (i+1, j-1) &\text{ if } D = L \\
        \end{cases}.
    \]
    We say that a supernode $S_{i, j}$ (resp. an entry $W_{i, j}$) is the \emph{parent} of another supernode $S_{i', j'}$ (resp. entry $W_{i', j'}$) in direction $D$ if the coordinate $(i, j)$ is the parent of $(i', j')$ in direction $D$.
\end{definition}

\noindent Each supernode may have up to four parents, depending on its position within the grid.
The parents in directions $V$, $R$, and $L$ enforce the transition consistencies for their child, as described in the previous section.
We also need a base case for the inductive process to work, which is to ensure that the LFMIS correctly recovers the first row of $W$.
We achieve this by enforcing a special set of horizontal consistencies for the supernodes on the first row through their parents in direction $H$.

For each direction \(D \in \cD\), we define a consistency function
\[
    \cC_D : \tiles  \to \cP(\tiles),
\]
where \(\cC_D\) maps each \(v \in \tiles\) to the subset of vertices \(v' \in \tiles\) such that, given the choice of $v$ for a supernode $S$, $v'$ is valid for the child supernode of $S$ in direction $D$ according to the machine's transition function.
In particular, the definition of $\cC_D$ is uniform in the sense that it is independent of the absolute location of $S$ within the grid.
We provide a formal definition for each $\cC_D$ below.

\paragraph*{Vertical Consistencies $\cC_V$} 
We give a detailed reasoning for the definition of $\cC_V$. $\cC_V$ describes the valid choices for an entry given the choice for its parent entry in the vertical direction $D$. Consider an input $(q, a) \in Q^* \times \Gamma$. 
\begin{itemize}
    \itemsep0em 
    \item If $q \in Q \setminus F$, it indicates that the tape head is over the cell represented by the parent entry in a non-halting state $q$. In this case, we have the full information about the child entry directly below, which represents the same tape cell but at the next time step: the Turing machine writes $\delta_{\Gamma} (q, a)$ at this tape cell and the tape head moves away.
    \item If $q \in F$ is one of the halting states, then the configuration of the Turing machine should not change and we simply copy $(q, a)$ for the vertical child entry.
    \item If $q = *$ is the dummy state, then we know that the alphabet $a$ must remain unchanged, but we do not have any information about the state. 
\end{itemize}

\noindent Formally, for each $(q, a) \in Q^* \times \Gamma$ we define
\[
   \cC_V(q, a) = 
    \begin{cases}
    \{(*, \delta _{\Gamma }(q, a))\}  & \text{ if } q \in Q \setminus F \\
    \{(q, a)\} &\text{ if } q \in F \\
    Q^* \times \{a\}  &\text{ if } q = *
    \end{cases}
\]
We complete the definition of $\cC_V$ by treating the special state $\hat{q}_0$ as $q_0$ and all special input alphabets $\hat{x}_i$ as $x_i$, and using the corresponding definitions above. 

\paragraph*{Horizontal Consistencies $\cC_H$} 
$\cC_H$ is defined primarily for the special symbols to ensure that the LFMIS correctly recovers the first row of the computation history $W$. 
The reason behind this will become clear in the proof later.
Define
 \[
   \cC_H(q, a) = 
    \begin{cases}
        Q^* \times (\Gamma \cup \{\hat{x}_1\}) & \text{ if } a = \hat{\$} \\
        Q^* \times (\Gamma \cup \{\hat{x}_{i+1}\}) &\text{ if } a =  \hat{x}_i, 1 \leq i \leq n - 1 \\
        Q^* \times \Gamma  &\text{ otherwise}
    \end{cases}
\]

\begin{remark}\label{remark:CH-contains-CV}
    We observe that $\cC_V(q, a) \subseteq Q^* \times \Gamma \subseteq \cC_H(q', a')$ for all $(q, a), (q', a') \in \tiles$.
\end{remark}

\begin{remark}
    Recall that the special input alphabets $\hat{x}_1, \hat{x}_2, \dotsc, \hat{x}_n$ are intentionally made distinct. This is to ensure that the second case of $\cC_H$ is well-defined.
\end{remark}

\paragraph*{Diagonally-Right Consistencies $\cC_R$}  
Consider an entry $w$ and its child entry $w'$ in the direction $R$.
$w'$ represents the tape cell right-adjacent to $w$, but at the next time step.
We observe that $w$ does not have any information about the alphabet of $w'$.
Moreover, the state of $w'$ is also unknown to $w$ unless the tape head is over the cell represented by $w$ in a non-halting state. So, for each $(q, a) \in Q^* \times \Gamma$, we define
\[
   \cC_R(q, a) = 
    \begin{cases}
       \{\delta_Q(q, a)\} \times \Gamma   &\text{ if \(q \in Q \setminus F\) and \(\delta_D(q, a) = R\)  }   \\
       \{*\} \times \Gamma   &\text{ if \(q \in Q \setminus F\) and \(\delta_D(q, a) = L\)  }   \\
        Q^* \times \Gamma   &\text{ if } q \in \{*\} \cup  F
    \end{cases},
\]
and treat the special symbols $\hat{q}_0$ as $q_0$, $\hat{\$}$ as $\$$, and each $\hat{x}_i$ as $x_i$.

\paragraph*{Diagonally-Left Consistencies $\cC_L$}
$\cC_L$ is symmetric to $\cC_R$, so for each $(q, a) \in Q^* \times \Gamma$, we define
\[
   \cC_L(q, a) = 
    \begin{cases}
       \{\delta_Q(q, a)\} \times \Gamma   &\text{ if \(q \in Q \setminus F \) and \(\delta_D(q, a) = L\)  }   \\
       \{*\} \times \Gamma   &\text{ if \(q \in Q \setminus F\) and \(\delta_D(q, a) = R\)  }   \\
        Q^* \times \Gamma   &\text{ if } q \in \{*\} \cup  F
    \end{cases},
\]
and treat the special symbols $\hat{q}_0$ as $q_0$, $\hat{\$}$ as $\$$, and each $\hat{x}_i$ as $x_i$.

We now proceed to the formal definition of $G$ and the proof of correctness.

\subsection{The Key Reduction to the LFMIS Problem}
\label{sec:LFMIS-reduction}

We present a generic reduction as mentioned in \Cref{lem:factored-LFMIS-reduction} from an arbitrary language $L \in \dtime(n^{\ell})$ to the LFMIS problem on factored graphs.
In this section, we give an explicit definition of the graph $G$ and show that the LFMIS of $G$ simulates the Turing machine deciding $L$.
In the next section, we give a factored graph representation for $G$, thus completing the reduction to the LFMIS problem on factored graphs.

Let \(L \in \dtime(n^{\ell})\) be a language and let $M$ be the Turing machine that decides $L$ within $O(n^{\ell})$ time.
There exists a constant $C > 0$ such that \(M\) halts within \(C \cdot n^{\ell}\) time for sufficiently large input size $n$.
Let $x = (x_1, \dotsc, x_n)$ be an input of size $n$.
Define $T$ to be the smallest integer power of $n$ such that
\[
    C \cdot n^{\ell} \leq T < nC \cdot n^{\ell},
\]
and let $\ell' := \log_n (T) = O(\ell)$ be the exponent.
We define a graph $G$ following the idea in \Cref{sec:computational-history-matrix}.

\subsubsection{Definition of Graph \texorpdfstring{$G$}{G}}
Recall that an instance of the LFMIS problem on factored graphs is a factored graph $G = f(G_1, \dotsc, G_k)$, with indices $V(G_i) = \{0, 1, \dotsc, |V(G_i)| - 1\}$ given for each graph factor $G_i$. The vertices in $G$ will be a flattened tuple of numbers. In this section, we define the explicit version $G = (V(G), E(G))$ of the factored graph $G$.

\paragraph*{Notation}
Let $b, k \geq 1$ be integers. For an integer $0 \leq y \leq b^k - 1$, we define the notation
\[
    [y]_b^k := (b_{k-1}, \dotsc, b_1, b_0) \in \Z_b^k
\]
to represent the $k$-bit base $b$ expansion of $y$, where $b_0$ is the least significant bit, so that $y = \sum_{i = 0}^{k-1} b_i b^i$.
For simplicity, the superscript $k$ may be omitted when the number of bits is clear from the context.

\paragraph*{Vertex Set}
The vertex set \(V(G)\) consists of \(T^2\) supernodes laid out in a \(T \times T\) grid structure.
Formally,
\[
 V(G) := \bigcup_{1 \leq i, j \leq T} S_{i, j},
\]
where $S_{i, j}$ denotes the supernode on the $i$-th row and $j$-th column of the grid, and is given by
\[
    S_{i, j} := \{([i - 1]_n^{\ell'}, [j - 1]_n^{\ell'}, q, a) \, \mid \, (q, a) \in \tiles\}.
\]
Unless otherwise stated, all the expansions used in this construction will be $\ell'$-bit base $n$ expansions, so we omit the superscript $\ell'$ for the remainder of this section.
For convenience, we also define the following notations for a vertex $v = ([i - 1]_n, [j - 1]_n, q, a) \in \Z_n^{\ell'} \times \Z_n^{\ell'} \times \tiles$ in $V(G)$:
\begin{align*}
    S(v) &: \textup{the supernode $S_{i, j}$ containing $v$} \\
    \cont(v) &: \textup{the state-alphabet pair, or content, $(q, a)$ represented by $v$} \\
    \state(v) &: \textup{the state $q$ represented by $v$} \\
    \alp(v) &: \textup{the alphabet $a$ represented by $v$}
\end{align*}

In order for $G = (V(G), E(G))$ to be the explicit version of a factored graph, the vertices need to be flattened tuples of numbers. The first $2\ell'$ coordinates of a vertex are already numbers, so it remains to provide indices for the states and alphabets represented by the vertices. For \(\hat{Q} = Q \cup \{*, \hat{q}_0\}\), we first map
\[
    \hat{q}_0 \mapsto 0, \qquad * \mapsto 1.
\]
The indices for the elements in $Q$ does not matter in the reduction, so we index $Q$ arbitrarily with $2, \dotsc, |\hat{Q}| - 1$. Similarly, for \(\hat{\Gamma} = \Gamma \cup \{\hat{x}_1, \dotsc, \hat{x}_n\}\), we map
\[
    \hat{\$} \mapsto 0, \quad \hat{x}_1 \mapsto 1, \quad \hat{x}_2 \mapsto 1, \quad \dotsc,  \quad \hat{x}_n \mapsto n, \quad \perp \, \mapsto n + 1,
\]
and arbitrarily index the elements in $\Gamma \setminus \{\perp\}$ with $n + 2, \dotsc, |\hat{\Gamma}| - 1$.

\paragraph*{Edge Set}
We break down the edge set $E(G)$ into the intra-supernode edges and inter-supernode edges:
\begin{itemize}
    \item {\bf Intra-supernode edges} \(E(G)_1\): We make a complete directed graph on each supernode. That is, we add bidirectional edges between all pairs of vertices in the same supernode,
    \[
        E(G)_1 := \{(v, v') \in V(G) \times V(G) \, \mid \,  S(v) = S(v')\}.
    \]
    \item {\bf Inter-supernode edges} \(E(G)_2\): Along each direction $D \in \cD$, we add directed edges from each parent supernode to their child supernode in direction $D$ according to the local consistencies defined in \Cref{sec:local-consistencies}: define
    \[
        E(G)_D := 
        \left\{ (v, v') \in V(G) \times V(G)  \, \middle\vert \,
            \begin{array}{c}
                \textup{$S(v)$ is a parent of $S(v')$ in direction $D$} \\
                \textup{and $\cont(v') \notin \cC_D(\cont(v))$}
            \end{array}
        \right\}.
    \]
    Then, we define the set of inter-supernode edges as the union 
    \[
        E(G)_2 := \bigcup_{D \in \cD} E(G)_D.
    \] 

\end{itemize}
Finally, the edge set \(E(G)\) of \(G\) is given by the union
\[
    E(G) := E(G)_1 \cup E(G)_2.
\]

This completes the definition of \(G = (V(G), E(G))\).

\subsubsection{Solving LFMIS on \texorpdfstring{$G$}{G} Simulates Turing Machine Computation}

In this section, we show that the explicit definition of graph $G$ is correct for the reduction. We state the correctness result as the following theorem.

\begin{theorem}\label{thm:LFMIS-simulates-TM-acceptance}
    The Turing machine \(M\) accepts on input \(x = (x_1, \dotsc , x_n )\) if and only if the vertex \
    \[
        s := ([T - 1]_n, [1]_n, \qacc, \perp) \in \lfmis(G).
    \]
\end{theorem}

Let $W$ be the $T \times T$ computational history matrix of $M$ on input $x$, as defined in \Cref{sec:computational-history-matrix}.
Recall that the first $2\ell'$ coordinates of $s$ indicates that it belongs to the supernode $S_{T, 2}$, which corresponds to the computational history matrix entry $W_{T, 2}$.
Based on the assumptions made about $M$ in \Cref{sec:Turing-machine-model}, $W_{T, 2}$ should be $(\qacc, \perp)$ if $M$ accepts or $(\qrej, \perp)$ otherwise.
Therefore, it suffices to show that $\lfmis(G)$ correctly \emph{recovers} the computational history $W$ of $M$ in order to prove \Cref{thm:LFMIS-simulates-TM-acceptance}.
This is interpreted in the following sense:

\begin{definition}
    For $1 \leq i \leq T$, we say that $\lfmis(G)$ \emph{recovers the $i$-th row of the computational history $W$} of the Turing machine $M$ on input $x$ if $\lfmis(G)$ contains exactly one vertex $v_{i, j}$ from each supernode $S_{i, j}$ on row $i$ such that
    \[
    \cont(v_{i, 1}), \cont(v_{i, 2}), \dotsc, \cont(v_{i, T})
    \]
    agrees with the $i$-th row of $W$.
\end{definition}

We use the rest of this section to prove that $\lfmis(G)$ recovers all rows of $W$, which proves \Cref{thm:LFMIS-simulates-TM-acceptance}.
Let $t \leq T$ be the \emph{exact} running time of $M$ on $x$; that is, $M$ halts exactly at time $t$.
We outline the proof as two steps: before the halting time $t$ and after $t$.
We first follow the idea in \Cref{sec:computational-history-matrix} and show that $\lfmis(G)$ recovers the first $t$ rows of $W$ based on the consistency functions defined for non-halting states in \Cref{sec:local-consistencies}.
If $t < T$, we need a second step to show that $\lfmis(G)$ correctly ``copies'' the vertex choices for the supernodes in any extra rows after the $t$-th row. This is achieved primarily through the vertical consistencies defined for halting states. 

We begin by noting the following properties of the graph $G$. The first property is immediate from the edges defined in $E(G)_1$:

\begin{proposition}\label{prop:at-most-one-each-supernode}
    Any independent set of \(G\) contains at most one vertex from each supernode.
\end{proposition}

The next lemma is the most important to this correctness proof, which characterizes the vertex that $\lfmis(G)$ contains from each supernode.
    
\begin{lemma}[Characterization of $\lfmis(G)$]\label{lem:characterization-of-LFMIS-G}
    Let \(S\) be a supernode in \(G\), and let \(P(S) \subseteq \cD\) be the subset of directions in which \(S\) has a parent supernode. Assume for each $D \in P(S)$, there is a vertex $v_D$ from the parent supernode of \(S\) in direction \(D\) that belongs to $\lfmis(G)$. Then, $\lfmis(G)$ contains the highest priority vertex $v \in S$ which satisfies the following intersection:
    \begin{equation}\label{eqn:consistency-intersection}
        \cont(v) \in \bigcap\limits_{D \in P(S)} \cC_D(\cont(v_D)) .
    \end{equation}
\end{lemma}
\begin{proof}
    This is a consequence of the characterization of the LFMIS for a general graph (see \Cref{prop:characterization-of-LFMIS}). 
    We first establish an order for the vertices in $G$.
    Recall that the LFMIS problem on factored graphs considers the vertices as they are ordered with respect to the lexicographic order on tuples $w = ([i - 1]_n, [j - 1]_n, q, a) \in \Z_n^{\ell'} \times \Z_n^{\ell'} \times \tiles$.
    Note that the lexicographic order on $\Z_n^{\ell'}$ agrees with the usual order on the decimal numbers they represent: $0, 1, \dotsc, n^{\ell'} - 1$.
    Therefore, the order of a vertex $w$ is given by considering 1) the row index $i$ of $S(w)$, 2) the column index $j$ of $S(w)$, 3) $\state(w)$, and 4) $\alp(w)$, in this order.

    We now start proving the lemma. We use the notations in the lemma statement, and let $u \in V(G)$ be a vertex in $\lfmis(G)$ with a higher priority than $v$. We consider two cases:

    \begin{itemize}
        \item {\bf Case 1.} $S(u) \neq S(v)$. 
        Then, according to the vertex orders, $S(u)$ must be from a row above or in the same row as $S(v)$ but to the left, 
        If $S(u)$ is not a parent of $S(v)$, then there is no edge between them.
        If $S(u)$ is a parent of $S(v)$, then $u$ must be one of the $v_D$'s. 
        But because $v$ satisfies \Cref{eqn:consistency-intersection}, in particular it satisfies $\cont (v) \in \cC_D(\cont(v_D)$, there is no edge between $v$ and $v_D (u = v_D)$ based on the definition of $E(G)_D$.

        \item {\bf Case 2.} $S(u) = S(v)$. Since $v$ has the highest priority among the vertices in $S(v)$ that satisfies \Cref{eqn:consistency-intersection}, it means $u \notin \cC_D(\cont(v_D))$ for some $D \in P(S)$. But we assumed $v_D \in \lfmis(G)$, so $u$ cannot belong to $\lfmis(G)$ and this case is not possible.
    \end{itemize}

    Therefore, there is no edge between $u$ and $v$, so we conclude that $v \in \lfmis(G)$ by \Cref{prop:characterization-of-LFMIS}.
\end{proof}

We now proceed to the first step of the proof.
\begin{lemma}\label{lem:LFMIS-recovers-first-t-rows}
    $\lfmis(G)$ recovers the first $t$ rows of $W$.
\end{lemma}  
\begin{proof}
    We follow the idea in \Cref{sec:computational-history-matrix} and prove by inducting on the rows. 

    \paragraph*{Base Case}
    We show that $\lfmis(G)$ recovers the first row of $W$. Recall that the first row of $W$ is 
    \[
        (\hat{q}_0, \hat{\$}), (*, \hat{x}_1), \dotsc, (*, \hat{x}_n), (*, \perp), \dotsc, (*, \perp).
    \]
     
    $S_{1, 1}$ does not have any parents, so \Cref{eqn:consistency-intersection} vacuously holds for all vertices in $S_{1, 1}$. Thus, $\lfmis(G)$ simply contains the highest priority vertex in $S_{1, 1}$, which is $([0]_n, [0]_n, \hat{q}_0, \hat{\$})$. Note that each remaining supernode on the first row only has the horizontal parent, so only $\cC_H$ matters in \Cref{eqn:consistency-intersection}.
    
    $S_{1, 2}$ has a horizontal parent $S_{1, 1}$ and $\cC_H(\hat{q}_0, \hat{\$}) = Q^* \times (\Gamma \cup \{\hat{x}_1\})$. The highest priority vertex in $S_{1, 2}$ satisfying this consistency is $([0]_n, [1]_n, *, \hat{x}_1)$.
    Inductively, for each $3 \leq i \leq n + 1$, if $\lfmis(G)$ contains the vertex $([0]_n, [i - 2]_n, *, \hat{x}_{i - 2}) \in S_{1, i - 1}$, then $\lfmis(G)$ contains $([0]_n, [i - 1]_n, *, \hat{x}_{i - 1}) \in S_{1, i}$, because it is the highest priority vertex in $S_{1, i}$ whose content satisfies $\cC_H(*, \hat{x}_{i - 2}) = Q^* \times (\Gamma \cup \{\hat{x}_{i - 1}\})$.

    Now, given that $\lfmis(G)$ contains $([0]_n, [n]_n, *, \hat{x}_n) \in S_{1, n + 1}$, with $\cC_H(*, \hat{x}_n) = Q^* \times \Gamma$, we know that $\lfmis(G)$ contains the vertex $([0]_n, [n + 1]_n, *, \perp) \in S_{1, n + 2}$. Another simple inductive argument shows that $\lfmis(G)$ contains the vertex $([0]_n, [i - 1]_n, * \perp) \in S_{1, i}$ for all $n + 3 \leq i \leq T$. Therefore, $\lfmis(G)$ recovers the first row of $W$.

    \bigskip

    There is a little caveat before proceeding to the inductive step.
    Recall from \Cref{sec:local-consistencies} that we replace the special symbols $\hat{q}_0$, $\hat{\$}$, and $\hat{x}_i$'s with the normal symbols they represent when they are inputs to $\cC_V$, $\cC_R$, or $\cC_L$. Since the supernodes in the first row of $G$ can only affect the choices of vertices in the second row through these consistencies, we may assume that the first row of $W$ is
    \[
    (q_0, \$), (*, x_1), \dotsc, (*, x_2), \dotsc, (*, x_n),
    \]
    and $\lfmis(G)$ contains the corresponding vertices from the first row.

    \paragraph*{Inductive Step}

    For the inductive hypothesis, assume that $\lfmis(G)$ recovers the $i$-th row of $W$ for some $1 \leq i < t$.
    We show that $\lfmis(G)$ recovers the $(i+1)$-th row of $W$ as well. We state two corollaries from \Cref{lem:characterization-of-LFMIS-G} that simplify the characterization of $\lfmis(G)$ for the supernodes below the first row.
    First, we observe that if a supernode is on the second row or below, then it always has a parent in the vertical direction. Thus, \Cref{remark:CH-contains-CV} immediately implies that the term $\cC_H(\cont(v_H))$ can be dropped from \Cref{eqn:consistency-intersection}:

    \begin{corollary}\label{cor:characterization-of-LFMIS-G-inductive}
        Assume the same preconditions as in \Cref{lem:characterization-of-LFMIS-G} and additionally assume that $S$ is below the first row. Then, $\lfmis(G)$ contains the highest priority vertex $v \in S$ which satisfies the following intersection:
        \begin{equation}\label{eqn:consistency-intersection-inductive}
        \cont(v) \in \bigcap\limits_{D \in P(S) \setminus \{H\}} \cC_D(\cont(v_D)) .
    \end{equation}
    \end{corollary}

\noindent For the other corollary, consider the special case where $\lfmis(G)$ contains the vertices representing the dummy state $*$ from all the parents in the row directly above (i.e. the parent in direction $V$ and/or $R, L$).
We show that $\lfmis(G)$ simply ``copies the content'' from the vertical parent in this case.

\begin{corollary}\label{cor:copy-content-if-all-star-states}
    Let \(S\) be a supernode in \(G\) below the first row.
    Assume that $\lfmis(G)$ contains the vertices representing state $*$ from all the parents of $S$ in the row directly above.
    If $\lfmis(G)$ contains the vertex $v'$ from the vertical parent of \(S\), then $\lfmis(G)$ will contain the vertex \(v \in S\) with \(\cont(v) = \cont(v')\).
\end{corollary}
\begin{proof}
    Say \(\cont(v') = (*, a')\). The vertical consistency requires \(\cC_V(*, a') = Q^* \times \{a'\}\).
    The diagonal consistencies \(\cC_R\) and \(\cC_L\) both map to \(Q^* \times \Gamma \supseteq Q^* \times \{a'\}\) whenever the state input is \(*\).
    Thus, the intersection in \Cref{eqn:consistency-intersection-inductive} always simplifies to \(Q^* \times \{a'\}\).
    Since we give \(*\) the highest priority in \(Q^*\), so $\lfmis(G)$ will contain the vertex representing \((*, a') = \cont(v')\).
\end{proof}
   
Now, to prove the inductive step, we make the following additional assumptions.
Notice that the supernodes on the left or right boundary of the grid only have two parents from the row directly above. 
In order to use the same argument for all the supernodes in a row, we may assume that the boundary supernodes \(S_{i+1, 1}\) 
and $S_{i+1, T}$ also have all three parents from the row directly above, for the following reasons.
We add an ``imaginary'' parent \(S_{i, 0}\) for \(S_{i+1, 1}\) and \(S_{i, T + 1}\) for \(S_{i+1, T}\), and assume that $\lfmis(G)$ contains the vertices representing $(*, \perp)$ from both imaginary parents.
Recall that \(S_{i+1, j}\) always has a vertical parent \(S_{i, j}\), and the vertical consistency \(\cC_V(q, a)\) only allows subsets of \(Q^* \times \Gamma \) for any choice of \((q, a) \in \tiles\).
Also recall that both diagonal consistencies \(\cC_R\) and \(\cC_L\) map to \(Q^* \times \Gamma\) whenever the state input is \(*\).
Therefore, the addition of imaginary parents doesn't change the intersection in \Cref{eqn:consistency-intersection-inductive}, hence not affecting the vertex choices in either boundary supernodes.

Let $S_{i+1, j}$, $1 \leq j \leq T$, be a supernode on the $(i+1)$-th row.
We consider different cases based on the vertices contained in $\lfmis(G)$ from the three parents \(S_{i, j-1}, S_{i, j}, S_{i, j+1}\).
For each \(j - 1 \leq  h \leq  j + 1\), say that $\lfmis(G)$ contains the vertex \(v_h = ([i - 1]_n, [h -1]_n, q_h, a_h) \in S_{i, h}\).
Note that since \(i < t\), each state $q_h$ must either be $*$ or a non-halting state in $Q \setminus F$.
Moreover, since the Turing machine only has a single tape head, at most one of \(q_{j-1}, q_j, q_{j+1}\) belongs to \(Q \setminus F\).
So we have the following four cases:    
\begin{enumerate}
    \item \(q_j \in Q \setminus F\) and \(q_{j-1} = q_{j+1} = *\). Let \((q^\prime , a_j^\prime , d) = \delta (q_j, a_j)\) denote the output of the transition function. Then, we know that the Turing machine \(M\) will overwrite the symbol at the \(j\)-th cell to \(a_j^\prime \) and move the tape head away from the \(j\)-th cell, so we must show that $\lfmis(G)$ contains the vertex \(([i]_n, [j - 1]_n, *, a_j^\prime )\) from \(S_{i + 1, j}\). We have
    \begin{align*}
        \cC_V(q_j, a_j) & = \{(*, a_j^\prime )\} \\
        \cC_R(q_{j-1} , a_{j-1} ) &= Q^* \times \Gamma \\
        \cC_L(q_{j+1} , a_{j+1} ) &= Q^* \times \Gamma 
    \end{align*}
    The intersection of the sets above is the singleton \(\{(*, a_j^\prime)\}\), so by \Cref{cor:characterization-of-LFMIS-G-inductive}, the $\lfmis(G)$ contains vertex \(([i]_n, [j - 1]_n, *, a_j^\prime)\) from \(S_{i+1, j}\).

    \item \(q_{j - 1} \in Q \setminus F\) and \(q_j = q_{j+1} = * \). Let \((q^\prime , a_{j-1} ^\prime , d) = \delta (q_{j-1} , a_{j-1} )\) denote the output of the transition function. Since the tape head is not over the \(j\)-th cell at time \(i\), the alphabet occupying the \(j\)-th cell should remain \(a_j\) at time \(i+1\). Moreover, the tape head will be pointing at the \(j\)-th cell in state \(q^\prime \) at time \(i+1\) if and only if \(d = R\). Thus, we must show that $\lfmis(G)$ contains the vertex \(([i]_n, [j - 1]_n, q^\prime, a_j)\) from \(S_{i+1, j}\) if \(d = R\), and \(([i]_n, [j - 1]_n, *, a_j)\) otherwise. We have 
    \begin{align*}
        \cC_V(q_j, a_j) &= Q^* \times \{a_j\} \\ 
        \cC_L(q_{j+1} , a_{j+1} ) &= Q^* \times \Gamma \\
        \cC_R(q_{j-1} , a_{j-1}) &=
        \begin{cases}
            \{q^\prime\} \times \Gamma & \text{ if } d = R\\
            \{*\} \times \Gamma & \text{ if } d = L
        \end{cases}
    \end{align*}
    
    The intersection of the sets above is the singleton \(\{(q^\prime , a_j)\}\) if $d = R$ and $\{(*, a_j)\}$ if $d = L$, so $\lfmis(G)$ contains the correct vertex in both cases.

    \item  \(q_{j + 1} \in Q \setminus F\) and \(q_{j-1} = q_j = *\). The reasoning is essentially the same as case (2) since the two diagonal consistencies \(\cC_L\) and \(\cC_R\) are symmetric. 
    
    \item \(q_{j - 1} = q_j = q_{j + 1} = *\). Again, since the tape head is not over the \(j\)-th cell at time \(i\), the alphabet occupying the \(j\)-th cell remains \(a_j\) at time \(i+1\). Now, notice that the tape head is only allowed to move one cell to the left or right at a time, so if the tape head is not over any of the \((j-1)\)-th, \(j\)-th, or \((j+1)\)-th cell at time \(i\), then it cannot be over the \(j\)-th cell at time \(i+1\). Thus, we must show that the LFMIS contains the node \(([i]_n, [j - 1]_n, *, a_j)\) from \(S_{i+1, j}\). But this is immediate from \Cref{cor:copy-content-if-all-star-states}.
\end{enumerate} 

We have shown that $\lfmis(G)$ contains the correct vertex from $S_{i+1, j}$ for all $1 \leq j \leq T$ on the $(i+1)$-th row, which completes the inductive step. The proof of the lemma is also complete by induction. 
\end{proof}

Now, if $t = T$ then we are done. Otherwise, we need to proceed to the second step in the proof outline.

\begin{lemma}\label{lem:LFMIS-recovers-extra-rows}
    If $t < T$, then $\lfmis(G)$ also recovers all the rows of $W$ below the $t$-th row.
\end{lemma}  
\begin{proof}
    Recall that the $t$-th row of $W$ represents the halting configuration of $M$, so if $t < T$, then all the rows of $W$ below the $t$-th row are identical to the $t$-th row.
    Note that the top-down edges between the supernodes in adjacent rows of $G$ are defined in the same way based on $\cC_V, \cC_R, \cC_L$, independent of the absolute row indices.
    Therefore, it suffices to show that $\lfmis(G)$ contains the vertices from each supernode in the $(t+1)$-th row that ``copies the content'' from the $t$-th row, then we can make the same conclusion for all remaining rows by a simple induction.

    The previous lemma \Cref{lem:LFMIS-recovers-first-t-rows} and the assumptions about $M$ in \Cref{sec:Turing-machine-model} indicate that $\lfmis(G)$ contains a vertex from each supernode in the $t$-th row, representing the following contents from $S_{t, 1}$ to $S_{t, T}$, respectively:
    \[
        (*, a_1), (q_F, a_2), (*, a_3), \dotsc, (*, a_T),
    \]
    where $a_1 a_2 \dotsc a_T$ is the string on the tape at time $t$, and $q_F = \qacc$ if $M$ accepts and $q_F = \qrej$ otherwise.
    
    We first note that \Cref{cor:copy-content-if-all-star-states} immediately implies that $\lfmis(G)$ contains the vertex representing $(*, a_j)$ from $S_{t+1, j}$ for all $4 \leq j \leq T$.
    For \(S_{t + 1, 1}\) and $S_{t + 1, 3}$, even though they have \(S_{t, 2}\) as one of their diagonal parents and $\lfmis(G)$ contains the vertex representing a halting state $q_F \in F$ from $S_{t+1, 2}$, both $\cC_{R / L} (*, a_2)$ and $\cC_{R / L}(q_F, a_2)$ map to \(Q^* \times \Gamma \).
    Thus, (only) for $S_{t+1, 1}$ and $S_{t+1, 3}$, we may assume that $\lfmis(G)$ contains the vertex $(*, a_2)$ from $S_{t, 2}$, so we can apply \Cref{cor:copy-content-if-all-star-states} again to get that $\lfmis(G)$ contains the vertices representing $(*, a_1), (*, a_3)$ from $S_{t+1, 1}, S_{t+1, 3}$, respectively.
    It remains to consider $S_{t+1, 2}$, and we use \Cref{cor:characterization-of-LFMIS-G-inductive}.
    We have
    \begin{align*}
        \cC_V(q_{F} , a_2) &= \{(q_F , a_2)\} \\ 
        \cC_R(* , a_1) &= Q^* \times \Gamma \\
        \cC_L(* , a_3) &= Q^* \times \Gamma
    \end{align*} 
    The intersection of the sets above is the singleton \(\{(q_F , a_2)\}\), so $\lfmis(G)$ contains the vertex representing $(q_F, a_2)$ from $S_{t+1, 2}$.
    This shows that $\lfmis(G)$ contains a vertex from each supernode in the $(t+1)$-th row representing the same content as those in the $t$-th row, which completes the proof.
\end{proof}

Combining \Cref{lem:LFMIS-recovers-first-t-rows} and \Cref{lem:LFMIS-recovers-extra-rows} proves \Cref{thm:LFMIS-simulates-TM-acceptance}.

\subsection{Factored Graph Construction of \texorpdfstring{$G$}{G} in the Reduction}
\label{sec:factored-graph-contruction-of-G}

    A naive direct representation of \(G\) has size \(\Omega (T^2) = \Omega(n^{2\ell})\), which is too much for the reduction to prove any meaningful lower bound for the LFMIS problem on factored graphs.
    Instead, we want to represent \(G\) as a factored graph, leveraging the fact that the computation rules of \(M\) are local (depending only on the symbol of the work tape under the head) and repetitive (the same rules apply regardless of the head's absolute position).
    In this section, we give a factored graph construction of the graph $G$ to complete the reduction. 
    Our strategy follows the outline in \Cref{sec:LFMIS-tech-overview} and \Cref{fig:LFMIS-factorization-roadmap}.
    Recall that $G$ is a grid of supernodes, and the edges $E(G)$ are partitioned into the set of intra-supernode edges $E(G)_1$ and inter-supernode edges $E(G)_2$. 
    We construct a factored graph representation for the subgraphs $G_1 := (V(G), E(G)_1)$ and $G_2 := (V(G), E(G)_2)$, respectively, and take a union in the end to recover $G$.

    Before describing the constructions, we introduce a simple yet powerful graph factor that will be helpful for the construction of both subgraphs.
    
    \begin{definition}[Empty Graph]
            An \emph{empty graph} on \(n\) vertices, denoted by \(\overline{K_n}\), is a graph with \(n\) vertices but no edges.  
    \end{definition}

    Empty graphs, when coupled with the Cartesian product, provides a very efficient way of duplication.
    
    \begin{lemma}\label{lem:empty-graph-product}
        For \(n, m \in \N\), we have
        \[
            \overline{K_n} \cart \overline{K_m} = \overline{K_{nm}}.  
        \] 
    \end{lemma}
    \begin{proof}
        \(\overline{K_n}\) and \(\overline{K_m}\) have \(n\) and \(m\) vertices respectively, so their product has \(nm\) vertices. Moreover, the product does not have any edges since neither of the factors does, so the equality holds.
    \end{proof}

    We define two duplication gadgets $E_T$ and $E_{T^2}$.
    Both gadgets are essentially empty graphs, with $T$ vertices and $T^2$ vertices, respectively.
    However, we give special names for the vertices, defined as $V(E_T) := \Z_n^{\ell'}$ and $V(E_{T^2}) := (\Z_n^{\ell'})^2$.
    It is not hard to see that the duplication gadgets themselves can be constructed efficiently using duplication.
    
    \begin{lemma}\label{lem:duplication-gadget}
        There exist factored graph representations of complexity $(n, O({\ell}))$ for both duplication gadgets, given by
        \[
            E_T = \underbrace{\overline{K_n} \cart \dotsb \cart \overline{K_n}}_{{\ell'} \textup{ times}} 
            \quad \textup{and} \quad 
            E_{T^2} = \underbrace{\overline{K_n} \cart \dotsb \cart \overline{K_n}}_{2{\ell'} \textup{ times}}.
        \]
    \end{lemma}
    \begin{proof}
        We need to first identify the vertex set of $\overline{K_n}$ as $\Z_n$. Then, \Cref{lem:empty-graph-product} together with a simple inductive argument shows that the above equalities hold. Clearly, both representations use $O(\ell') = O(\ell)$ input graphs of size \(n\).  
    \end{proof}

    \begin{remark}
        Since empty graphs have no edges, the spatial locations of their vertices are not specified.
        However, it would be helpful to visualize the vertices of $E_T$ as arranged in a line, and the vertices of $E_{T^2}$ as arranged in a $T \times T$ grid.
    \end{remark}

    Another important point to note is that the size of each supernode is linear in $n$.

    \begin{proposition}\label{prop:supernode-linear-size}
            The set $\dot{Q} \times \dot{\Gamma}$ has size \(O(n)\).  
    \end{proposition} 
    \begin{proof}
        This is because in the set 
        \[\dot{Q} \times \dot{\Gamma} = \dot{Q} \times (\Gamma \cup \{\dot{x} _1, \dotsc , \dot{x} _n \}),\]
        the only part that is dependent on the input size \(n\) is \(\{\dot{x} _1, \dotsc , \dot{x} _n \}\).
        The sizes of \(\Gamma \) and \(\dot{Q} = Q \cup \{*, \dot{q} _0\}\) are dependent only on the definition of the Turing Machine \(M\), but are independent of the input size \(n\). 
        So the total size of $\dot{Q} \times \dot{\Gamma}$ is \(O(n)\).
    \end{proof}

    We now proceed to the factored graph construction for both subgraphs.

    \subsubsection{Recovering the Intra-supernode Edges}\label{sec:intra-supernode edges}

    We start by recovering the subgraph $G_1$ with all the intra-supernode edges.
    To be precise, we give a factored graph representation for the subgraph $G_1 = (V(G), E(G)_1)$, which is a grid of complete supernodes.
    
    Let \(K\) be the complete digraph on the vertices \(\dot{Q} \times \dot{\Gamma}\).
    Then, we can easily duplicate $K$ into a $T \times T$ grid of $K$'s with the duplication gadget $E_{T^2}$.
    
    \begin{lemma}\label{lem:supergrid with E1}
        There exists a factored graph representation of complexity $(O(n), O(\ell))$ for the subgraph $G_1$ given by 
        \[
            G_1 = E_{T^2} \cart K.
        \]
    \end{lemma}
    \begin{proof}
        Denote the product by $P := E_{T^2} \cart K$.
        It is clear that the vertex set of $P$ is $\Z_n^{\ell'} \times \Z_n^{\ell'} \times \tiles$, which is equal to $V(G)$.
        Since \(E_{T^2}\) does not have any edges, we can only apply one of the edge conditions in Cartesian product: for two vertices \(v = (\mathbf{i}, \mathbf{j}, q, a), v' = (\mathbf{i'}, \mathbf{j'}, q^\prime , a^\prime ) \in \Z_n^{\ell'} \times \Z_n^{\ell'} \times \tiles\) in \(P\), there is a directed edge \((v, v')\) in \(P\) if and only if $(\mathbf{i}, \mathbf{j}) = (\mathbf{i'}, \mathbf{j'})$ and $((q, a), (q', a')) \in E(K)$. 
        The first condition means that $S(v) = S(v')$, and the second condition is always true since \(K\) is the complete directed graph on \(\dot{Q}  \times \dot{\Gamma}\).
        Therefore, $P$ has exactly the same set of edges as \(E(G)_1\). The complexity follows from \Cref{lem:duplication-gadget} and \Cref{prop:supernode-linear-size}.       
    \end{proof}

    \subsubsection{Recovering the Inter-supernode Edges}\label{sec:inter-supernode edges}

    It now remains to give a factored graph representation for the subgraph $G_2 = (V(G), E(G)_2)$ with all the inter-supernode edges.
    Recovering the inter-supernode edges turns out to be a lot more complicated than recovering the intra-supernode edges, and the process will be broken down into several stages.
    We follow a bottom-up approach according to the outline in \Cref{fig:LFMIS-factorization-roadmap}, starting from the simpler graph gadgets.
    We first give an efficient factored graph representation for paths.
    Then, for each direction $D \in \cD$, we combine paths and/or empty graphs in some way to construct a grid $G_D$.
    $G_D$ consists of normal nodes instead of supernodes, and only grid points neighboring in the direction $D$ are connected by edges.
    We use tensor product to ``embed'' the consistency relation described by $\cC_D$ into the corresponding grid $G_D$, which will give the graph $(V(G), E(G)_D)$.
    Finally, taking the union of the factored graph representations for $(V(G), E(G)_D)$ over $D \in \cD$ recovers all the inter-supernode edges.

    \paragraph*{Paths} 

    We show that if the length of a path is a perfect power $b^k$, then the path has a factored graph representation using $k^2$ input graphs of size $b$. Note that the number of input graphs is only poly-logarithmic in the length of the path.
        
    \begin{lemma}\label{lem:inc path}
        For all integers $b, k > 1$, there exists a factored graph representation of complexity $(b, k^2)$ for the path of length $b^k$.
    \end{lemma}
    
    \begin{proof}
        Let \(b, k\) be given.
        We label the vertices of the path with $k$-bit base $b$ expansions in increasing order of $0, 1, \dotsc, b^k - 1$ along the path, so we write the path as
        \[
            \pi := ([0]_b^k, [1]_b^k, \dotsc, [b^k - 1]_b^k).
        \]
        The intuition of the factored graph construction is then to think of each edge as representing the ``plus one'' operation from one of its endpoints to the other.
        Observe that for each $0 \leq y < b^k - 1$, when adding one to its $k$-bit base $b$ expansion $[y]_b^k : = b_{k-1} \dotsc b_1 b_0$, there exists an index $0 \leq i \leq k - 1$ such that
        \begin{enumerate}
            \item the $i$ least significant bits $b_{i-1}, \dotsc, b_1, b_0$ are wrapped around from $b - 1$ to 0
            \item the bit $b_i$ is incremented by 1
            \item all other higher bits $b_{k-1} ,\dotsc, b_{i+1}$ remain unchanged.
        \end{enumerate}
        We partition the edges based on the number of least significant bits being wrapped around from $b - 1$ to 0, and give a factored graph representation for each subgraph induced by the partitions. For each \(0 \leq i \leq k - 1\), define a graph $\pi_i = (V(\pi_i), E(\pi_i))$, where \(V(\pi_i) = V(\pi)\) and 
        \[
            E(\pi_i) := \left\{ ([y]_b^k, [y+1]_b^k) \in E(\pi)  \middle\vert 
            \begin{array}{c}
                \textup{$[y+1]_b^k$ wraps around exactly the $i$ least} \\
                \textup{significant bits of $[y]_b^k$ from $b - 1$ to 0}
            \end{array}
            \right\}.
        \]
        Notice that \(\pi_0, \dotsc , \pi_{k-1}\) partition the edges \(E(\pi)\) of the path, so we can simply use union operations to recover
        \[
            \pi = \pi_0 \cup  \pi_1 \cup \dotsb  \cup  \pi_{k-1}. 
        \]
        
        \begin{figure}[!ht]
    \centering
    \tikzstyle{state} = [circle, draw, minimum size=25pt, inner sep=0pt]
        \begin{tikzpicture}[auto, node distance=2.5cm, thick, scale=1, every node/.style={scale=0.4}]
            % Diagram 1
            \begin{scope}[shift={(-20cm, 0cm)}]
                \tikzset{vertex/.style={circle, draw, fill=black, inner sep=0pt, minimum size=5pt}}

                \colorlet{p0}{darkgray}
                \colorlet{p1}{purple}
                \colorlet{p2}{pink}
        
                % Vertices
                \node[vertex] (A1) at (0,0) {};
                \node[vertex] (A2) at (1,0) {};
                \node[vertex] (A3) at (2,0) {};
                \node[vertex] (B1) at (0,-1) {};
                \node[vertex] (B2) at (1,-1) {};
                \node[vertex] (B3) at (2,-1) {};
                \node[vertex] (C1) at (0,-2) {};
                \node[vertex] (C2) at (1,-2) {};
                \node[vertex] (C3) at (2,-2) {};
        
                \node[vertex] (D1) at (4,0) {};
                \node[vertex] (D2) at (5,0) {};
                \node[vertex] (D3) at (6,0) {};
                \node[vertex] (E1) at (4,-1) {};
                \node[vertex] (E2) at (5,-1) {};
                \node[vertex] (E3) at (6,-1) {};
                \node[vertex] (F1) at (4,-2) {};
                \node[vertex] (F2) at (5,-2) {};
                \node[vertex] (F3) at (6,-2) {};
        
                \node[vertex] (G1) at (8,0) {};
                \node[vertex] (G2) at (9,0) {};
                \node[vertex] (G3) at (10,0) {};
                \node[vertex] (H1) at (8,-1) {};
                \node[vertex] (H2) at (9,-1) {};
                \node[vertex] (H3) at (10,-1) {};
                \node[vertex] (I1) at (8,-2) {};
                \node[vertex] (I2) at (9,-2) {};
                \node[vertex] (I3) at (10,-2) {};

                % Numbers below vertices
                \tikzset{label/.style={yshift=-0.5cm, scale=2}}
                \node[label] at (A1.south)  {000};
                \node[label] at (A2.south)  {001};
                \node[label] at (A3.south)  {002};
                \node[label] at (B1.south)  {010};
                \node[label] at (B2.south)  {011};
                \node[label] at (B3.south)  {012};
                \node[label] at (C1.south)  {020};
                \node[label] at (C2.south)  {021};
                \node[label] at (C3.south)  {022};
        
                \node[label] at (D1.south)  {100};
                \node[label] at (D2.south)  {101};
                \node[label] at (D3.south)  {102};
                \node[label] at (E1.south)  {110};
                \node[label] at (E2.south)  {111};
                \node[label] at (E3.south)  {112};
                \node[label] at (F1.south)  {120};
                \node[label] at (F2.south)  {121};
                \node[label] at (F3.south)  {122};
        
                \node[label] at (G1.south)  {200};
                \node[label] at (G2.south)  {201};
                \node[label] at (G3.south)  {202};
                \node[label] at (H1.south)  {210};
                \node[label] at (H2.south)  {211};
                \node[label] at (H3.south)  {212};
                \node[label] at (I1.south)  {220};
                \node[label] at (I2.south)  {221};
                \node[label] at (I3.south)  {222};

                \path (A1) edge [->, very thick, color=p0] (A2)
                      (A2) edge [->, very thick, color=p0] (A3)
                      (A3) edge [->, very thick, color=p1] (B1)
                      (B1) edge [->, very thick, color=p0] (B2)
                      (B2) edge [->, very thick, color=p0] (B3)
                      (B3) edge [->, very thick, color=p1] (C1)
                      (C1) edge [->, very thick, color=p0] (C2)
                      (C2) edge [->, very thick, color=p0] (C3)
                      (C3) edge [->, very thick, color=p2] (D1);
            
                \path (D1) edge [->, very thick, color=p0] (D2)
                      (D2) edge [->, very thick, color=p0] (D3)
                      (D3) edge [->, very thick, color=p1] (E1)
                      (E1) edge [->, very thick, color=p0] (E2)
                      (E2) edge [->, very thick, color=p0] (E3)
                      (E3) edge [->, very thick, color=p1] (F1)
                      (F1) edge [->, very thick, color=p0] (F2)
                      (F2) edge [->, very thick, color=p0] (F3)
                      (F3) edge [->, very thick, color=p2] (G1);
        
                \path (G1) edge [->, very thick, color=p0] (G2)
                      (G2) edge [->, very thick, color=p0] (G3)
                      (G3) edge [->, very thick, color=p1] (H1)
                      (H1) edge [->, very thick, color=p0] (H2)
                      (H2) edge [->, very thick, color=p0] (H3)
                      (H3) edge [->, very thick, color=p1] (I1)
                      (I1) edge [->, very thick, color=p0] (I2)
                      (I2) edge [->, very thick, color=p0] (I3);
                  
            \node at (12,-1) [draw=none, fill=none, scale=2] {
                \begin{tabular}{@{}l@{}l}
                    \textcolor{p0}{Black edges: } & $\pi_0$ \\
                    \textcolor{p1}{Red edges: } & $\pi_1$ \\
                    \textcolor{p2}{Pink edges: } & $\pi_2$ \\
                \end{tabular}
            };
            \end{scope}
        \end{tikzpicture}
    \caption{Example partition for $\pi$ where $b = 3, k = 3$. The black edges belong to $\pi_0$, red edges belong to $\pi_1$, and pink edges belong to $\pi_2$.}
    \label{fig:path-decomp}
\end{figure}
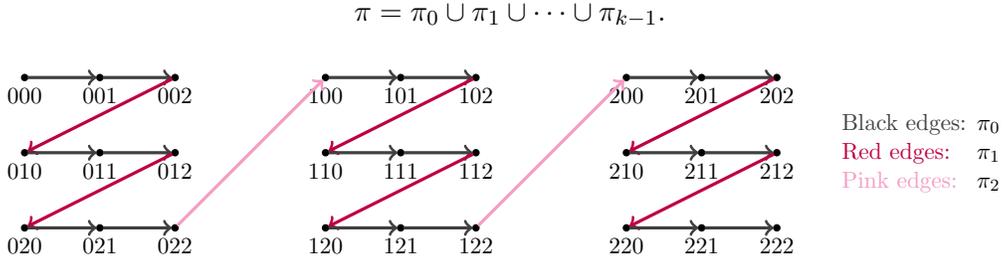
        
        We now factorize the $\pi_i$'s to achieve the desired complexity. Define three graph factors \(A, B,\) and \(C\), where all three graphs have the same vertex set \(V_b := \{0, 1, \dotsc , b-1 \}\) and 
        \begin{enumerate}
            \item \(A\) does not have any edges
            \item \(B\) has a directed edge from \(i\) to \(i + 1\) for all \(0 \leq  i < b - 1\)
            \item \(C\) has a single directed edge from \(b - 1\) to 0.
        \end{enumerate}

        \begin{figure}[!ht]
    \centering
    \tikzstyle{state} = [circle, draw, minimum size=25pt, inner sep=0pt]

    \begin{tikzpicture}[auto, node distance=2.5cm, thick, scale=0.5, every node/.style={scale=0.5}]
         % Labels A, B, C
        \tikzset{heading/.style={yshift=1, scale=2}}
        \node[heading] at (-20cm, 1cm) {$A$};
        \node[heading] at (-10cm, 1cm) {$B$};
        \node[heading] at (0cm, 1cm) {$C$};
        
        % Diagram A
        \begin{scope}[shift={(-20cm, 0cm)}]
            \tikzset{vertex/.style={circle, draw, fill=black, inner sep=0pt, minimum size=5pt}}

            % Vertices
            \node[vertex] (A1) at (0,0) {};
            \node[vertex] (A2) at (2,0) {};
            \node[label, scale=2]  (A4) at (4,0) {$\dotsb$};
            \node[vertex] (A5) at (6,0) {};

            \tikzset{label/.style={yshift=-0.4cm, scale=2}}
            \node[label] at (A1.south)  {0};
            \node[label] at (A2.south)  {1};
            \node[label] at (A5.south)  {$b-1$};
        \end{scope}
    
        % Diagram B
        \begin{scope}[shift={(-10cm, 0cm)}]
            \tikzset{vertex/.style={circle, draw, fill=black, inner sep=0pt, minimum size=5pt}}
            % Vertices
            \node[vertex] (A1) at (0,0) {};
            \node[vertex] (A2) at (2,0) {};
            \node[label, scale=2]  (A4) at (4,0) {$\dotsb$};
            \node[vertex] (A5) at (6,0) {};

            \tikzset{label/.style={yshift=-0.4cm, scale=2}}
            \node[label] at (A1.south)  {0};
            \node[label] at (A2.south)  {1};
            \node[label] at (A5.south)  {$b-1$};

            \path (A1) edge [->, thick] (A2)
                  (A2) edge [->, thick] (A4)
                  (A4) edge [->, thick] (A5);
        \end{scope}
    
        % Diagram C
        \begin{scope}[shift={(0cm, 0cm)}]
            % Vertices
            \tikzset{vertex/.style={circle, draw, fill=black, inner sep=0pt, minimum size=5pt}}
            % Vertices
            \node[vertex] (A1) at (0,0) {};
            \node[vertex] (A2) at (2,0) {};
            \node[label, scale=2]  (A4) at (4,0) {$\dotsb$};
            \node[vertex] (A5) at (6,0) {};

            \tikzset{label/.style={yshift=-0.4cm, scale=2}}
            \node[label] at (A1.south)  {0};
            \node[label] at (A2.south)  {1};
            \node[label] at (A5.south)  {$b-1$};

            \path (A5) edge [bend right, ->, thick] (A1);
        \end{scope}
    \end{tikzpicture}
    \caption{Graph Factors $A$, $B$, and $C$.}
    \label{fig:path-construction-graph-factor}
\end{figure}
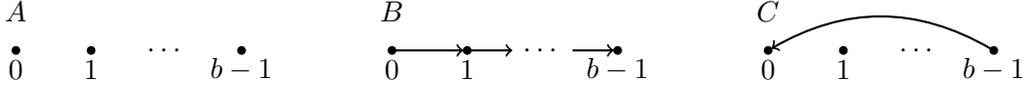

        One can intuitively think of the vertex set $V_b$ as representing the base $b$ digits, \(A\) as representing the unchanged digits, \(B\) as representing the digit being incremented by 1, and \(C\) as representing the \(i\) least significant digits being wrapped around. Following this intuition, we make the following claim.
        
        \begin{claim}\label{claim:factored path}
             For each \(0 \leq  i \leq  k - 1\), \(\pi_i\) has a factored graph representation of complexity $(b, k)$ given by 
             \[
                \pi_i = \underbrace{(A \cart \dotsb \cart A)}_{k - i - 1 \textup{ times}} \cart \, (B \times \underbrace{C \times \dotsb \times C}_{i \textup{ times} }).
             \]   
        \end{claim}

        \begin{proof}[Proof of \Cref{claim:factored path}]
        Denote the product by $P_i := A^{\cart (k-i-1)} \cart (B \times C^{\times i})$. 
        $V(P_i)$ clearly equals \((V_b)^k = V(\pi)\) from the definition, so it remains to show that $E(P_i) = E(\pi_i).$
        Let \((b_{k-1}, \dotsc , b_1, b_0)\) and \((c_{k-1}, \dotsc , c_1, c_0)\) be two vertices in $P_i$.
        We first unpack the middle Cartesian product separating the two parentheses.
        Note that \(A\) does not have any edges, so neither does any Cartesian product \(A \cart \dotsb \cart A\).
        Thus, only one of the edge conditions in the Cartesian product can be applied: there is a directed edge \(((b_{k-1}, \dotsc , b_1, b_0), (c_{k-1}, \dotsc , c_1, c_0))\) in $P_i$ if and only if \((b_{k-1}, \dotsc , b_{i+1}) = (c_{k-1},\dotsc , c_{i+1})\) and \(((b_i, \dotsc , b_1, b_0), (c_i, \dotsc , c_1, c_0)) \in E(B \times C^{\times i})\).
        If we further unpack the tensor product between \(B\) and \(C^{\times i}\), this edge exists if and only if
            \begin{itemize}
                \item \((b_i ,c_i) \in E(B)\), so we have \(0 \leq  b_i < b - 1\) and \(b_i + 1= c_i\), and 
                \item  \((b_j, c_j) \in E(C)\) for all \(0 \leq  j \leq  i - 1\), so we have \(b_j = b - 1\) and \(c_j = 0\).  
            \end{itemize}
        Therefore, this is exactly the same set of edges as \(E(\pi_i)\). This representation clearly uses $k$ input graphs of size $b$.
        \end{proof}

        It follows from \Cref{claim:factored path} that \(\pi_0 \cup \pi_1 \cup  \dotsb \cup \pi_{k-1}\) is indeed a factored graph representation of \(\pi\), and it uses $k^2$ input graphs of size $b$.
    \end{proof}
    
     We summarize the first important gadget in the factored graph construction of $G$ as the following lemma.
    \begin{lemma}\label{lem:path-complexity}
        The increasing path (resp. decreasing path)
        \[
            \vec{\pi}_T := ([0]_n, [1]_n, \dotsc , [T-1]_n) \quad \textup{(resp. \( \cev{\pi}_T := ([T-1]_n, \dotsc , [1]_n, [0]_n)\))} 
        \]
        has a factored graph representation of complexity $(n, O(\ell^2)).$
    \end{lemma} 
    \begin{proof}
        The factored graph representation for the increasing path $\vec{\pi}_T$ is immediate from (the proof of) \Cref{lem:inc path} if we set \(b = n\) and \(k = \ell' = O(\ell)\).
        Similarly, we obtain a factored graph representation of the same complexity for the decreasing path $\cev{\pi}_T$ with the exact same construction, except that we reverse all the edges in the graph factors $B$ and $C$.
    \end{proof}

    \paragraph*{Grids with Edges in a Single Direction}\label{subsec:grids with consistency}
    
    Recall from \Cref{sec:LFMIS-reduction} that \(G\) is a \(T \times T\) grid of supernodes. There are edge connections that represent the consistencies between pairs of nodes, where each node is from a different supernode and the supernodes are neighbors in one of the directions in $\cD = \{V, H, R, L\}$. In this section, for each direction $D \in \cD$, we give a factored graph representation for the \(T \times T\) grid \(G_D\) of normal nodes (instead of supernodes), with edges connecting all pairs of nodes neighboring in the direction \(D\).

    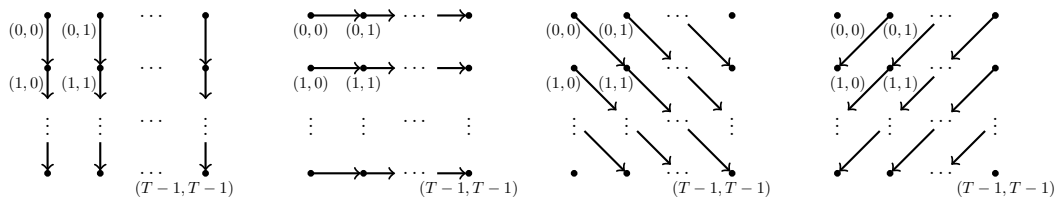
\begin{figure}[ht]
            \centering
            \tikzstyle{state} = [circle, draw, minimum size=25pt, inner sep=0pt]
                
                \begin{tikzpicture}[auto, node distance=2.5cm, thick, scale=0.7, every node/.style={scale=0.35}]
                    % Diagram 1
                    \begin{scope}[shift={(-20cm, 0cm)}]
                        \tikzset{vertex/.style={circle, draw, fill=black, inner sep=0pt, minimum size=5pt}}
            
                        % Vertices
                        \node[vertex] (A1) at (0,0) {};
                        \node[vertex] (A2) at (1,0) {};
                        \node[scale=2] (A4) at (2,0) {$\dotsc$};
                        \node[vertex] (A5) at (3,0) {};
                    
                        \node[vertex] (B1) at (0,-1) {};
                        \node[vertex] (B2) at (1,-1) {};
                        \node[scale=2] (B4) at (2,-1) {$\dotsc$};
                        \node[vertex] (B5) at (3,-1) {};
        
                        \node[scale=2]  (D1) at (0,-2) {$\vdots$};
                        \node[scale=2]  (D2) at (1,-2) {$\vdots$};
                        \node[scale=2] (D4) at (2,-2) {$\dotsc$};
                        \node[scale=2]  (D5) at (3,-2) {$\vdots$};
                        
                        \node[vertex] (E1) at (0,-3) {};
                        \node[vertex] (E2) at (1,-3) {};
                        \node[scale=2] (E4) at (2,-3) {$\dotsc$};
                        \node[vertex] (E5) at (3,-3) {};

                        % Numbers below vertices
                        \tikzset{label/.style={yshift=-0.5cm, xshift=-0.8cm, scale=2}}
                        \node[label, scale=0.8] at (A1.south)  {$(0, 0)$};
                        \node[label, scale=0.8] at (A2.south)  {$(0, 1)$};

                        \node[label, scale=0.8] at (B1.south)  {$(1, 0)$};
                        \node[label, scale=0.8] at (B2.south)  {$(1, 1)$};
                        
                        \node[label, scale=0.8] at (E5.south)  {$(T-1, T-1)$};
                        
                        \path 
                              (A1) edge [->] (B1)
                              (A2) edge [->] (B2)
                              (A5) edge [->] (B5)
                              (B1) edge [->] (D1)
                              (B2) edge [->] (D2)
                              (B5) edge [->] (D5)
                              (D1) edge [->] (E1)
                              (D2) edge [->] (E2)
                              (D5) edge [->] (E5);
                    \end{scope}

                    % Diagram 2
                    \begin{scope}[shift={(-15cm, 0cm)}]
                        \tikzset{vertex/.style={circle, draw, fill=black, inner sep=0pt, minimum size=5pt}}
            
                        % Vertices
                        \node[vertex] (A1) at (0,0) {};
                        \node[vertex] (A2) at (1,0) {};
                        \node[scale=2] (A4) at (2,0) {$\dotsc$};
                        \node[vertex] (A5) at (3,0) {};
                    
                        \node[vertex] (B1) at (0,-1) {};
                        \node[vertex] (B2) at (1,-1) {};
                        \node[scale=2] (B4) at (2,-1) {$\dotsc$};
                        \node[vertex] (B5) at (3,-1) {};
        
                        \node[scale=2]  (D1) at (0,-2) {$\vdots$};
                        \node[scale=2]  (D2) at (1,-2) {$\vdots$};
                        \node[scale=2] (D4) at (2,-2) {$\dotsc$};
                        \node[scale=2]  (D5) at (3,-2) {$\vdots$};
                        
                        \node[vertex] (E1) at (0,-3) {};
                        \node[vertex] (E2) at (1,-3) {};
                        \node[scale=2] (E4) at (2,-3) {$\dotsc$};
                        \node[vertex] (E5) at (3,-3) {};

                        % Numbers below vertices
                        \tikzset{label/.style={yshift=-0.5cm, scale=2}}
                        \node[label, scale=0.8] at (A1.south)  {$(0, 0)$};
                        \node[label, scale=0.8] at (A2.south)  {$(0, 1)$};

                        \node[label, scale=0.8] at (B1.south)  {$(1, 0)$};
                        \node[label, scale=0.8] at (B2.south)  {$(1, 1)$};
                        
                        \node[label, scale=0.8] at (E5.south)  {$(T-1, T-1)$};

                        \path
                              (A1) edge [->] (A2)
                              (A2) edge [->] (A4)
                              (A4) edge [->] (A5)
                              (B1) edge [->] (B2)
                              (B2) edge [->] (B4)
                              (B4) edge [->] (B5)
                              (E1) edge [->] (E2)
                              (E2) edge [->] (E4)
                              (E4) edge [->] (E5);
                    \end{scope}

                    % Diagram 3
                    \begin{scope}[shift={(-10cm, 0cm)}]
                        \tikzset{vertex/.style={circle, draw, fill=black, inner sep=0pt, minimum size=5pt}}
            
                        % Vertices
                        \node[vertex] (A1) at (0,0) {};
                        \node[vertex] (A2) at (1,0) {};
                        \node[scale=2] (A4) at (2,0) {$\dotsc$};
                        \node[vertex] (A5) at (3,0) {};
                    
                        \node[vertex] (B1) at (0,-1) {};
                        \node[vertex] (B2) at (1,-1) {};
                        \node[scale=2] (B4) at (2,-1) {$\dotsc$};
                        \node[vertex] (B5) at (3,-1) {};
        
                        \node[scale=2]  (D1) at (0,-2) {$\vdots$};
                        \node[scale=2]  (D2) at (1,-2) {$\vdots$};
                        \node[scale=2] (D4) at (2,-2) {$\dotsc$};
                        \node[scale=2]  (D5) at (3,-2) {$\vdots$};
                        
                        \node[vertex] (E1) at (0,-3) {};
                        \node[vertex] (E2) at (1,-3) {};
                        \node[scale=2] (E4) at (2,-3) {$\dotsc$};
                        \node[vertex] (E5) at (3,-3) {};

                        % Numbers below vertices
                        \tikzset{label/.style={yshift=-0.5cm, xshift=-0.4cm, scale=2}}
                        \node[label, scale=0.8] at (A1.south)  {$(0, 0)$};
                        \node[label, scale=0.8] at (A2.south)  {$(0, 1)$};

                        \node[label, scale=0.8] at (B1.south)  {$(1, 0)$};
                        \node[label, scale=0.8] at (B2.south)  {$(1, 1)$};
                        
                        \node[label, scale=0.8] at (E5.south)  {$(T-1, T-1)$};
                        \path 
                              (A1) edge [->] (B2)
                              (A2) edge [->] (B4)
                              (A4) edge [->] (B5)
                              (B1) edge [->] (D2)
                              (B2) edge [->] (D4)
                              (B4) edge [->] (D5)
                              (D1) edge [->] (E2)
                              (D2) edge [->] (E4)
                              (D4) edge [->] (E5);
                              ;
                    \end{scope}

                    \begin{scope}[shift={(-5cm, 0cm)}]

                        \tikzset{vertex/.style={circle, draw, fill=black, inner sep=0pt, minimum size=5pt}}
            
                        % Vertices
                        \node[vertex] (A1) at (0,0) {};
                        \node[vertex] (A2) at (1,0) {};
                        \node[scale=2] (A4) at (2,0) {$\dotsc$};
                        \node[vertex] (A5) at (3,0) {};
                    
                        \node[vertex] (B1) at (0,-1) {};
                        \node[vertex] (B2) at (1,-1) {};
                        \node[scale=2] (B4) at (2,-1) {$\dotsc$};
                        \node[vertex] (B5) at (3,-1) {};
        
                        \node[scale=2]  (D1) at (0,-2) {$\vdots$};
                        \node[scale=2]  (D2) at (1,-2) {$\vdots$};
                        \node[scale=2] (D4) at (2,-2) {$\dotsc$};
                        \node[scale=2]  (D5) at (3,-2) {$\vdots$};
                        
                        \node[vertex] (E1) at (0,-3) {};
                        \node[vertex] (E2) at (1,-3) {};
                        \node[scale=2] (E4) at (2,-3) {$\dotsc$};
                        \node[vertex] (E5) at (3,-3) {};

                        % Numbers below vertices
                        \tikzset{label/.style={yshift=-0.5cm, xshift=0.4cm, scale=2}}
                        \node[label, scale=0.8] at (A1.south)  {$(0, 0)$};
                        \node[label, scale=0.8] at (A2.south)  {$(0, 1)$};

                        \node[label, scale=0.8] at (B1.south)  {$(1, 0)$};
                        \node[label, scale=0.8] at (B2.south)  {$(1, 1)$};
                        
                        \node[label, scale=0.8] at (E5.south)  {$(T-1, T-1)$};
                        
                        \path 
                              (A2) edge [->] (B1)
                              (A4) edge [->] (B2)
                              
                              (A5) edge [->] (B4)
                              (B2) edge [->] (D1)
                              (B4) edge [->] (D2)
                              
                              (B5) edge [->] (D4)
                              (D2) edge [->] (E1)
                              (D4) edge [->] (E2)
                              
                              (D5) edge [->] (E4);
                    \end{scope}
                
                \end{tikzpicture}
            \caption{Grids $G_V, G_H, G_R,$ and $G_L$ (from left to right).}
            \label{fig:grids}
        \end{figure}
            
    We now give a formal definition of each grid $G_D$. The vertex set is $(\Z_n^{\ell'})^2$. Here, we use a pair of vectors \((\mathbf{i}, \mathbf{j}) \in (\Z_n^{\ell'})^2\) to represent the coordinate of each vertex in the grid, where \(\mathbf{i} = [i]_n\) and $\mathbf{j} = [j]_n$ are the ${\ell'}$-bit base $n$ expansions of the row index $i$ and column index $j$, respectively.
    For simplicity, we may interchangeably use the pair of decimal indices $(i, j) \in (\Z_T)^2$ to represent the same vertex.
    For two vertices \((i, j), (i^\prime , j^\prime ) \in (\Z_T)^2\) in \(G_D\), there is a directed edge \(((i, j),(i^\prime , j^\prime ))\) if and only if $(i, j)$ is a parent of $(i', j')$ in direction $D$ (see \Cref{def:parent}).

    We start with the vertical and horizontal grids. Intuitively, we only need to duplicate a path \(T\) times to get $G_V$ and $G_H$, and duplication can be achieved efficiently with Cartesian products and empty graphs.

    \begin{lemma}\label{lem:vertical grid}
        The vertical grid \(G_V\) and the horizontal grid $G_H$ both have factored graph representations of complexity $(O(n), O(\ell^2))$ given by
        \begin{align*}
            G_V &= \vec{\pi}_T \cart E_T\\
            G_H &= E_T \cart \vec{\pi}_T
        \end{align*}
    \end{lemma}
    \begin{proof}
         We prove the equation for \(G_V\), and a similar argument works for \(G_H\).
         Denote the product by $P := \vec{\pi}_T \cart E_T$. 
         Both factors \(\vec{\pi }_T\) and \(E_T \) have \(\Z_n^{\ell'}\) as the vertex set, so the vertex set of $P$ is \((\Z_n^{\ell'})^2\).
         It remains to show that $P$ has the same set of edges as \(G_V\).
         Similar to $G_V$, we use a pair of decimal indices \((i, j)\) to represent each vertex in the product \(P\).
         Let \((i, j)\) and \((i^\prime , j^\prime )\) be two vertices in \(P\).
         Since \(E_T\) does not have any edges, there is only one way to unpack the Cartesian product: there is a directed edge \(((i, j), (i^\prime , j^\prime ))\) in \(P\) if and only if $j' = j$ and $(i, i') \in E(\vec{\pi}_T)$, meaning $i' = i + 1$.
         This describes the exact same set of edges as in \(G_V\). The complexity follows from \Cref{lem:path-complexity} and \Cref{lem:duplication-gadget}.
    \end{proof}

    Next, we build the diagonal grids $G_R$ and $G_L$. We use the right-diagonal grid $G_R$ as an example to illustrate the intuition. Observe that the edge condition
    \[
    i' = i + 1 \quad \textup{and} \quad j' = j + 1
    \]
    looks like a conjunction of two edge conditions of paths. This motivates the use of tensor product on two increasing paths.
    
    \begin{lemma}\label{lem:diagonal grid}
        The diagonal grids \(G_R\) and \(G_L\) both have factored graph representations of complexity $(O(n), O({\ell}^2))$ given by
        \begin{align*}
            G_R &= \vec{\pi}_T \times \vec{\pi}_T\\
            G_L &= \vec{\pi}_T \times \cev{\pi}_T
        \end{align*}
    \end{lemma}
    \begin{proof}
        We prove the equation for \(G_R\), and a similar argument works for \(G_L\).
        Again, it is clear that the vertex set of the product \(\vec{\pi}_T \times \vec{\pi}_T\) is $(\Z_n^{\ell'})^2$, so it remains to show that the edge sets are equal. 
        We also use the decimal indices \((i, j)\) to represent each vertex in the product \(\vec{\pi}_T \times \vec{\pi}_T\). 
        Let \((i, j)\) and \((i^\prime , j^\prime)\) be two vertices in \(\vec{\pi}_T \times \vec{\pi}_T\). 
        By the definition of tensor product, there is an edge \(((i, j), (i^\prime , j^\prime ))\) in \(\vec{\pi}_T \times \vec{\pi}_T\) if and only if $(i, i') \in E(\vec{\pi}_T)$ and $(j, j') \in E(\vec{\pi}_T)$, which means $i' = i + 1$ and $j' = j + 1$, respectively.
        This is exactly the same set of edges as in \(G_R\), so the equation \(G_R = \vec{\pi}_T \times \vec{\pi}_T\) holds. 
        The complexity follows from \Cref{lem:path-complexity}.
    \end{proof}

    \paragraph*{Grids of Supernodes with Consistency Connections}
    In this section, we apply the tensor product to \emph{embed} consistency relations into the grids $G_D$ that we constructed previously. 
    This transforms each grid into a grid of supernodes, with edges going between different supernodes encoding consistency relations. 

    Let us begin with a simple example to illustrate the idea of embedding. 
    Let $L$ be a line segment (or, a path of length one) and $T$ be a triangle. 
    Then, $L \times T$ is a bipartite graph, where a copy of $V(T)$ is placed at each endpoint of the line segment $L$, corresponding to the parts of the bipartite graph, and the edges go across different copies of $T$ following the relation described by $E(T).$
    
    \begin{figure}[!ht]
    \centering
    \begin{tikzpicture}[auto, node distance=1.5cm, thick, scale=0.9]
        % Left Side
        % Top Graph (2x2 Grid with directed edges)
        \begin{scope}[xshift=-6cm, yshift=1cm]
            \tikzset{vertex/.style={circle, draw, fill=black, inner sep=0pt, minimum size=5pt}}
    
            % Vertices
            \node[vertex] (A1) at (0,0) {};
            \node[vertex] (A2) at (2,0) {};
    
            \tikzset{label/.style={yshift=0.4cm, scale=1}}
            \node[label] at (A1.south)  {$0$};
            \node[label] at (A2.south)  {$1$};

            \path 
                  (A1) edge [->, very thick] (A2);

            \node at (-0.5, 1) {\textbf{$L$}};
          
        \end{scope}

        % Bottom Graph (Complete Graph K5)
        \begin{scope}[xshift=-5.9cm, yshift=-2cm]
            \tikzset{vertex/.style={circle, draw, fill=black, inner sep=0pt, minimum size=5pt}}
    
            % Vertices
            \node[vertex] (A1) at (0,-0.5) {};
            \node[vertex] (A2) at (2,-0.5) {};
            \node[vertex] (A3) at (1,1) {};
    
            \tikzset{label/.style={yshift=-0.3cm, scale=1}}
            \node[label] at (A1.south)  {$a$};
            \node[label] at (A2.south)  {$b$};

            \tikzset{label/.style={yshift=0.4cm, scale=1}}
            \node[label] at (A3.south)  {$c$};

            \path 
                  (A1) edge [->, very thick] (A2)
                  (A2) edge [->, very thick] (A3)
                  (A3) edge [->, very thick] (A1);

            \node at (-0.5, 1) {\textbf{$T$}};

        \end{scope}

        % Tensor Product Symbol between A and B
        \node at (-5cm, 0cm) {\textbf{\LARGE $\times$}};

        \node at (-2.7cm, 0cm) {\textbf{\LARGE $\Rightarrow$}};

        % Right Side (Grid of K5 Complete Graphs)
        \begin{scope}[xshift=0cm, yshift=-1cm]
            \def\n{3} % Number of nodes in each complete graph
            \def\radius{.7cm} % Radius of the circle for the graph
            \def\circleRadius{1.2cm} % Radius of the dashed circle around the graph
            \def\spacing{5cm} % Spacing between graphs in the grid
        
            \foreach \i in {0,1} {
                \begin{scope}[xshift=\i*\spacing]
                    % Draw a dashed circle around the complete graph
                    \draw[dashed] (0, 0.2) circle (\circleRadius);
                    
                    % Place the nodes in a circular arrangement with unique names
                    \foreach \k in {1,...,\n} {
                        \node[circle, fill, inner sep=2pt, label={$\alphalph{\k}$}] (v\i\k) at 
                        ({\radius * cos(360/\n * \k)}, {\radius * sin(360/\n * \k)}) {};
                    }
                
                    % Note: the edges should also reference the correct unique node names
                    
                \end{scope}
            }

            \node at (-\spacing/4, \circleRadius + 0.5) {$0$}; % Label for \i = 0
            \node at (\spacing/1.33, \circleRadius + 0.5) {$1$}; % Label for \i = 1
            
            % Now you can draw edges between specific nodes
            \draw[->, very thick, bend left=30] (v02) to[out=-45, in=-135] (v13);
            \draw[->, very thick, bend right=30] (v01) to[out=45, in=135] (v12);
            \draw[->,very thick, bend left=30] (v03) -- (v11);

            \node at (-1, 2) {\textbf{$L \times T$}};

        \end{scope}
        
    \end{tikzpicture}
    \caption{Example of Embedding a Triangle Relation into a Line Segment}
    \label{fig:tensor-product-ex}
\end{figure}
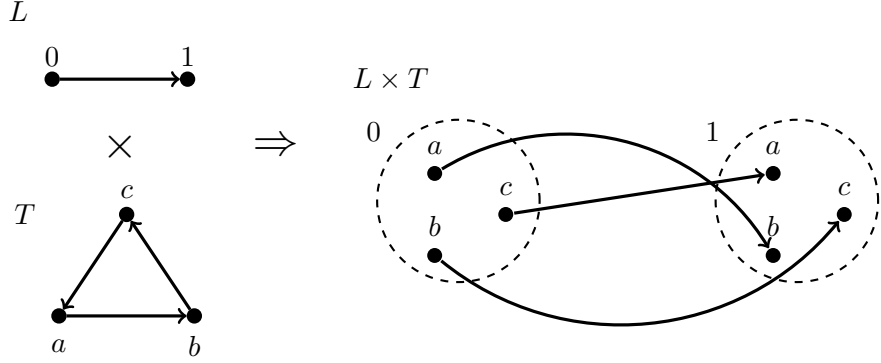
    
    Therefore, a nice way to think of the tensor product is as follows: if $G$ and $H$ are arbitrary graphs, then the tensor product \(G \times H\) can be viewed as embedding the ``adjacency relation'' of $H$ into the ``structure'' of $G$. In our instance, we define a \emph{consistency relation graph} $R_D$ that encodes the consistency function $\cC_D$, and embed the relation of $R_D$ into the structure of $G_D$. For each direction \(D \in \cD\), we define $R_D$ as follows. The vertex set $V(R_D) = \tiles$, where the indices for $\dot{Q}$ and $\dot{\Gamma}$ are defined according to the orderings in \Cref{sec:LFMIS-reduction}. For any two vertices \((q, a), (q^\prime , a^\prime )\) of \(R_D\) (not necessarily distinct), there is a directed edge \(((q, a), (q^\prime , a^\prime ))\) in $R_D$ if and only if \((q^\prime , a^\prime ) \notin \cC_D(q, a)\). First notice that \(R_D\) has size \(O(n)\).

    We can then embed \(R_D\) into the grid \(G_D\) that we constructed previously.

    \begin{figure}[!ht]
    \centering
    \begin{tikzpicture}[auto, node distance=1.5cm, thick]
        % Left Side
        % Top Graph (2x2 Grid with directed edges)
        \begin{scope}[xshift=-5.5cm, yshift=1cm]
            \tikzset{vertex/.style={circle, draw, fill=black, inner sep=0pt, minimum size=5pt}}
    
            % Vertices
            \node[vertex] (A1) at (0,0) {};
            \node[vertex] (A2) at (1,0) {};
            \node[vertex] (B1) at (0,-1) {};
            \node[vertex] (B2) at (1,-1) {};
    
            \tikzset{label/.style={yshift=0.4cm, scale=1}}
            \node[label] at (A1.south)  {(0, 0)};
            \node[label] at (A2.south)  {(0, 1)};

            \tikzset{label/.style={yshift=-0.3cm, scale=1}}
            \node[label] at (B1.south)  {(1, 0)};
            \node[label] at (B2.south)  {(1, 1)};
            
            \path 
                  (A1) edge [->, very thick] (B1)
                  (A2) edge [->, very thick] (B2);

            \node at (-0.5, 1) {\textbf{$G_V$}};
          
        \end{scope}

        % Bottom Graph (Complete Graph K5)
        \begin{scope}[xshift=-5cm, yshift=-3cm]
            % Parameters
            \def\n{5}  % Number of nodes in the complete graph
            \def\radius{1cm} % Radius of the circle for the graph
        
            % Place the nodes in a circular arrangement
            \foreach \i in {1,...,\n} {
                \node[circle, fill, inner sep=2pt, label={$\alphalph{\i}$}] (v\i) at 
                ({\radius * cos(360/\n * \i)}, {\radius * sin(360/\n * \i)}) {};
            }
        
            % Draw edges between all pairs of nodes
            \draw[->, very thick] (v4) -- (v5);
            \path[->, very thick] (v2) edge[out=30, in=-30, looseness=20] (v2);

             \node at (-1, 1.5) {\textbf{$R_V$}};

        \end{scope}

        % Tensor Product Symbol between A and B
        \node at (-5cm, -1cm) {\textbf{\LARGE $\times$}};

        \node at (-2.7cm, -1cm) {\textbf{\LARGE $\Rightarrow$}};

        % Right Side (Grid of K5 Complete Graphs)
        \begin{scope}[xshift=0cm]
            \def\n{5} % Number of nodes in each complete graph
            \def\radius{.7cm} % Radius of the circle for the graph
            \def\circleRadius{1.2cm} % Radius of the dashed circle around the graph
            \def\spacing{2.8cm} % Spacing between graphs in the grid
        
            \foreach \i in {0,1} {
                \foreach \j in {0,1} {
                    % Position each graph in the grid
                    \begin{scope}[xshift=\i*\spacing, yshift=-\j*\spacing]
                        % Draw a dashed circle around the complete graph
                        \draw[dashed] (0, 0.2) circle (\circleRadius);
                        
                        % Place the nodes in a circular arrangement with unique names
                        \foreach \k in {1,...,\n} {
                            \node[circle, fill, inner sep=2pt, label={$\alphalph{\k}$}] (v\i\j\k) at 
                            ({\radius * cos(360/\n * \k)}, {\radius * sin(360/\n * \k)}) {};
                        }
                    
                        % Note: the edges should also reference the correct unique node names
                        
                    \end{scope}
                }
            }

            % Place labels for the graphs
            \node at (-\spacing/2, \circleRadius + 0.5) {$(0,0)$}; % Label for the top-left graph
            \node at (\spacing/2, \circleRadius + 0.5) {$(0,1)$}; % Label for the top-right graph
            \node at (-\spacing/2, -\circleRadius*3 - 0.5) {$(1,0)$}; % Label for the bottom-left graph
            \node at (\spacing/2, -\circleRadius*3 - 0.5) {$(1,1)$}; % Label for the bottom-right graph
            
            % Now you can draw edges between specific nodes
            \draw[->, very thick, bend left=30] (v002) to[out=-45, in=-135] (v012);
            \draw[->, very thick, bend right=30] (v004) to[out=45, in=135] (v015);
            
            \draw[->,very thick, bend left=30] (v102) to[out=-45, in=-135] (v112);
            
            \draw[->, very thick, bend right=30] (v104) to[out=45, in=135] (v115);

            \node at (-1, 2) {\textbf{$G_V \times R_V$}};

        \end{scope}
        
    \end{tikzpicture}
    \caption{Example of Embedding the Vertical Relation $R_V$ into the Vertical Grid $G_V$}
    \label{fig:tensor-product-LFMIS}
\end{figure}
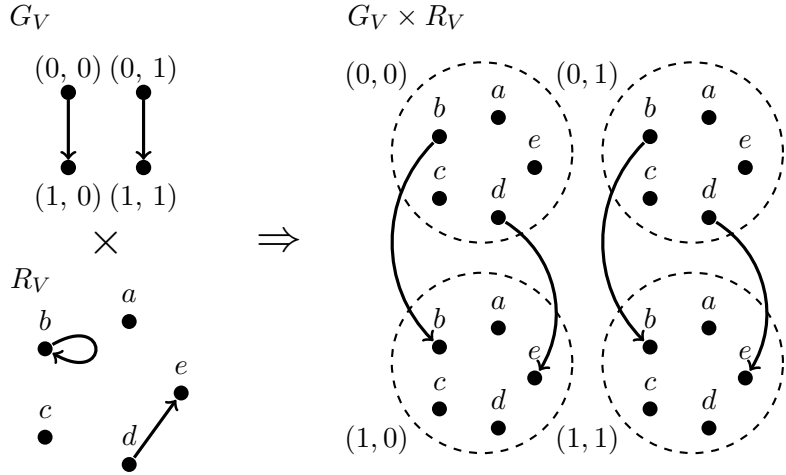
    
    \begin{lemma}\label{lem:embed consistency}
        For each direction \(D \in \cD\), there is a factored graph representation of complexity $(O(n), O(\ell^2))$ for the graph $(V(G), E(G)_D)$ given by
        \[
            G_D \times R_D.
        \]
    \end{lemma}
    \begin{proof}
        Denote the product by $P := G_D \times R_D$.
        It is clear that the vertex set of $P$ is $\Z_n^{\ell'} \times \Z_n^{\ell'} \times \tiles$, which is equal to $V(G)$.
        For two vertices \(v = (\mathbf{i}, \mathbf{j}, q, a), v' = (\mathbf{i'}, \mathbf{j'}, q^\prime , a^\prime ) \in \Z_n^{\ell'} \times \Z_n^{\ell'} \times \tiles\) in \(P\), by the definition of tensor product, there is a directed edge \((v, v')\) in \(P\) if and only if \(((\mathbf{i}, \mathbf{j}), (\mathbf{i'}, \mathbf{j'})) \in E(G_D)\) and \(((q, a), (q', a')) \in E(R_D)\).
        The first condition means that the supernode $S(v)$ is a parent of $S(v')$ in direction \(D\), and the second condition means that \((q', a') \notin \cC_D(q, a)\).
        These are exactly the same conditions as in the definition of \(E(G)_D\), so the equality holds. The complexity follows from \Cref{lem:vertical grid}, \Cref{lem:diagonal grid}, and \Cref{prop:supernode-linear-size}.
    \end{proof}

    Now, it immediately follows from \Cref{lem:embed consistency} and the definition of \(E(G)_2\) that we recover all the inter-supernode edges $E(G)_2$ by simply taking the union of $G_D \times R_D$ over all $D \in \cD$.
    
    \begin{lemma}\label{lem:supergrid with E2}
    There exists a factored graph representation of complexity $(O(n), O(\ell^2))$ for the subgraph $G_2$ given by
    \[
        G_2 = \bigcup_{D \in \cD} G_D \times R_D.
    \]
    \end{lemma}

    \subsubsection{Summary}
    To conclude this section, we take the union of the two subgraphs $G_1$ (\Cref{lem:supergrid with E1}) and $G_2$ (\Cref{lem:supergrid with E2}) to complete the factorization of $G$.
    
    \begin{theorem}\label{thm:LFMIS-factored-complexity}
        There is a factored graph representation of complexity $(O(n), O(\ell^2))$ for the graph \(G = (V(G), E(G))\) given by
        \[
            G = G_1 \cup G_2.
        \]
        In particular, constructing this representation takes $O(\ell^2 n^2)$ time.
    \end{theorem}
    \begin{proof}
        The correctness and complexity of the representation follows from \Cref{lem:supergrid with E1}, \Cref{lem:supergrid with E2}, and the definition that $E(G) = E(G)_1 \cup E(G)_2$.
        To construct this representation, we generate $O(\ell^2)$ input graphs, each of size at most $O(n)$.
        Since constructing each input graph takes at most quadratic time in its size, the overall construction time is $O(\ell^2 n^2)$.
    \end{proof}
    
    Now, note that \Cref{lem:factored-LFMIS-reduction} follows from the correctness in \Cref{thm:LFMIS-simulates-TM-acceptance} and the complexity in \Cref{thm:LFMIS-factored-complexity}.
    This completes the proof of \Cref{thm:factored-LFMIS-XP-complete}.

\section{Counting Small Cliques on Factored Graphs}
\label{sec:counting-subgraphs}

We prove that counting constant-sized cliques is fixed-parameter tractable on factored graphs.
In this section, we assume that all graphs are undirected.

\begin{definition}
    \label{def:exact-copies}
    Suppose $H$ is a graph on $s$ vertices $b_1, \dotsc, b_{s}$.
    Let $a_{1}, \dotsc, a_{s}$ be an (ordered) collection of $s$ distinct vertices in a graph $G$.
    $a_{1}, \dotsc, a_{s}$ form an \emph{exact copy} of $H$ if for all $i, j \in [s]$, $a_{i}, a_{j}$ are adjacent if and only if $b_{i}, b_{j}$ are adjacent.
\end{definition}

\CountingSubgraphs*

Consider an $s$-tuple of vertices in $G$, denotes $v_{1}, \dotsc, v_{s}$.
While the $s$-tuple may form a clique in a single factored component $G_{F}$, it may also do so using edges from different factored components to make up the clique.
We define a decomposition of the clique $H$ as a map from the edges of $H$ to the factored components.
Thus, we count the number of $s$-cliques satisfying the decomposition, that is, each edge exists in the assigned factored component.
However, this may lead to some double counting, so that we apply the Inclusion-Exclusion principle and instead count the number of $s$-cliques that satisfy any arbitrary collection $\cC$ of decompositions.

Thus, fix a collection $\cC$ of decompositions.
We partition all $s$-tuples into $2^{|\cC| s^{2} k}$ classes, where each class is determined by whether $v_{i}, v_{j}$ satisfy the relevant adjacency and equality conditions for a given coordinate $\ell \in [k]$ and decomposition in $\cC$.
Note that we obtain this upper bound on the number of classes since for each decomposition in $\cC$ and for each of the $\binom{s}{2}$ pairs of vertices, there are at most $2k$ binary conditions to verify. 
We observe that either every $s$-tuple in a class satisfies every decomposition in the collection or none do.
Thus, it suffices to count the number of $s$-tuples in each class and sum up the relevant classes (i.e., the ones where the class satisfies every decomposition in the collection).
To compute this count efficiently, we observe that whether the adjacency and equality conditions are met in one coordinate does not affect the conditions in other coordinates.
Thus, it suffices to compute the number of satisfying $s$-tuples in each coordinate separately and then obtain the overall count via a product.
In each coordinate, the graph size is at most $|\cC| s^{2} n$, since for each decomposition in the collection $|\cC|$ there are at most $s^{2}$ different factored components.
Therefore, we can count the number of satisfying tuples in each coordinate in $|\cC|^{s} s^{2s} n^{s}$ time, which is in $\fpt$.

\begin{proof}[Proof of \Cref{thm:counting-subgraphs}]
    We count the exact copies of $H$ in $G$ as follows.
    For any dimension $d \leq k$, let $V_{d}$ denote the set of vertices with dimension $d$.
    Note $k$ is the maximum dimension since any factored component has tree structure with at most $k$ leaves.
    
    First, \Cref{lemma:factored-component-same-dimension} shows that there are no edges between vertices of distinct dimensions.
    In particular, we can compute the number of exact copies on each separate component and sum to obtain
    \begin{equation*}
        \countH(G) = \sum_{d} \countH(G[V_{d}])
    \end{equation*}

    where the sum is over the distinct vertex dimensions of $V(G)$.
    The following lemma then describes how efficiently we can compute $\countH(G[V_{d}])$ for each dimension $d$.
    
    \begin{lemma}[Counting Edges for a Dimension]
        \label{lemma:num-edges-one-type}
        For each dimension $d$, there is an algorithm computing $\countH(G[V_{d}])$ in $O(g'(s, k) n^{s})$-time for some function $g'(s, k)$.
    \end{lemma}

    \begin{proof}[Proof of \Cref{lemma:num-edges-one-type}]
        We begin by bounding the number of factored components of a given dimension.
    
        \begin{lemma}
            \label{lemma:num-factored-components}
            There are at most $2^{k}$ distinct factored components with dimension $d$.
        \end{lemma}

        \begin{proof}[Proof of \Cref{lemma:num-factored-components}]
            An internal node $v$ in the tree structure of $G$ with operation $\cup$ and $c_{v} \geq 2$ children gives $c_{v}$ possible choices for choosing a factored component.
            On the other hand, for an internal node $v$ to have $c_{v}$ children, it must correspond to $c_{v} - 1$ operations in the factored graph formula.
            Therefore, the number of factored components is at most the product of $c_{v}$ over all internal nodes $v$ with operation $\cup$.
            In particular, we have the quantity
            \begin{equation*}
                \prod_{v} c_{v}
            \end{equation*}
            under the constraint $\sum_{v} (c_{v} - 1) = u$, or equivalently $u + n_{u} = \sum_{v} c_{v}$, where $u$ is the number of union operations and $n_{u} \leq u$ is the number of internal nodes with operation $\cup$ in the tree structure of $G$.
            Therefore, we have $\sum_{v} c_{v} \leq 2u$ and thus the total number of distinct factored components is at most $2^{u} \leq 2^{k - 1} < 2^{k}$.
        \end{proof}

        An exact copy of $H$ in $V_{d}$ may not entirely lie in a single factored component $G_{F}$.
        Instead, an exact copy of $H$ may have different edges being drawn from different factored components.
        We define a {\bf decomposition} of $H$ as an arbitrary partition of the edges of $H$ into factored components.
        In particular, a decomposition of $H$ assigns a factored component to each edge $(h_{i}, h_{j})$.
        Since there are $\binom{s}{2}$ pairs of vertices in $H$ and at most $2^{k}$ factored components, there are at most $c(s, k) = \binom{s}{2}^{2^{k}}$ decompositions of $H$.

        We describe how to compute the number of exact copies of a given decomposition of $H$.
        We say that an $s$-tuple of vertices in $G$ satisfies the decomposition if $(v_{i}, v_{j}) \in E(G_{F(i, j)})$ for all $i \neq j$, where $G_{F(i, j)}$ is the factored component assigned to $(h_{i}, h_{j})$.
        Note that if an $s$-tuple satisfies the decomposition, this $s$-tuple forms an $s$-clique in $G$ and thus contributes to $\countH(G)$.
        Furthermore, any exact copy of $H$ must satisfy some decomposition.
        Of course, this may lead to double counting since a single $s$-tuple may satisfy multiple decompositions.
        To remedy this, we use the Inclusion-Exclusion principle.
        In particular, for any collection $\cC$ of decompositions (note that there are at most $2^{c(s, k)}$ such collections), we count the number of $s$-tuples satisfying every decomposition in this collection.

        Suppose now we are given a collection $\cC$ of decompositions.
        Note that $\cC$ has at most $c(s, k)$ decompositions.
        We categorize all $s$-tuples of vertices in $V_{d}$ according to their adjacency and identity relations in the input graphs.
        For each factored component $G_{F}$, let $G_{F, \ell}$ denote the $\ell$-th input graph.
        Formally, we construct a table with $2^{|\cC| \binom{s}{2} 2 d}$ rows and $|\cC| \binom{s}{2} 2 d$ columns, where $s$ is the number of vertices in $H$. 
        Denote an arbitrary $s$-tuple of vertices as $v_{1}, \dotsc, v_{s}$.
        We denote the $d$ coordinates of $v_{i}$ with $v_{i, \ell}$ for $\ell \in [d]$, so $v_{i} = (v_{i, 1}, \dotsc, v_{i, d})$.
        The columns encode the following conditions for every decomposition in $\cC$.
        \begin{enumerate}
            \item $v_{i, \ell} = v_{j, \ell}$ where $i \neq j$ and $\ell \in [d]$
            \item $(v_{i, \ell}, v_{j, \ell}) \in E(G_{\ell})$ where $i \neq j$ and $G_{\ell}$ is the $\ell$-th input graph in the factored component assigned to $(h_{i}, h_{j})$.
        \end{enumerate}
        Thus, there are at most $|\cC| 2 \binom{s}{2} k$ columns.
        We build a truth table over these columns, thus obtaining a table with $2^{|\cC| 2 \binom{s}{2} k}$ rows.
        We say that an $s$-tuple \emph{satisfies} some row if the $s$-tuple agrees with the conditions imposed by the row of the truth table.
        We now argue that 
        \begin{enumerate}
            \item each $s$-tuple of vertices in $G_{F}$ satisfies exactly one row
            \label{item:rows-partition-tuples}
            
            \item for each row, either every $s$-tuple satisfying this row satisfies every decomposition in $\cC$ or none of them do.
            \label{item:rows-determine-satisfication}
        \end{enumerate}

        The first claim follows since the conditions of each row are mutually exclusive and since we include all possible conditions, each $s$-tuple must satisfy at least one (and therefore exactly) one row.
        To verify the second claim, consider a $s$-tuple of vertices.
        We note that they satisfy all decompositions in $\cC$ if and only if for every decomposition $(v_{i}, v_{j}) \in E(G_{F})$ for the factored component $G_{F}$ assigned to $(h_{i}, h_{j})$.
        To verify that $(v_{i}, v_{j}) \in E(G_{F})$, it suffices to check the following:
        \begin{enumerate}
            \item If $v_{i}$ or $v_{j}$ are not in $V(G_{F})$, then clearly there is no edge in $G_{F}$. 

            \item If $v_{i}, v_{j}$ are in $V(G_{F})$, then whether $(v_{i}, v_{j}) \in E(G_{F})$ can be checked by examining the tree structure of the factored component $G_{F}$ and the input graphs labeling the leaves of $G_{F}$.
            Note that at each node, the verification (either in tensor or Cartesian product nodes) consists only of checks for equality and/or adjacency.
        \end{enumerate}

        Our above discussion leads to the following algorithm for computing the number of $s$-tuples satisfying every decomposition in $\cC$.
        
        \begin{enumerate}
            \item  {\bf Classifying tuples.}
            After creating the table for the collection $\cC$, we iterate through each row of the table. 
            Each row represents the set of $s$-tuples of vertices that satisfy the row $r$.

            Above we have shown that every $s$-tuple satisfying row $r$ satisfies every decomposition or none of them do.
            To check which is the case, we simply check if for every decomposition in $\cC$, the $s$-tuple forms an $s$-clique in $O(|\cC| s^2 k)$ time (since checking one edge in one decomposition requires $O(k)$ time).
            
            Let $N_{r}$ denote the number of $s$-tuples that satisfy row $r$.
            We add $N_{r}$ to the overall count if and only if the $s$-tuples satisfying row $r$ satisfy every decomposition in $\cC$.
            Computing the set of rows for which we would like to sum up $N_{r}$ requires $O(2^{|\cC| s^2 k} |\cC| s^2 k)$ time.
    
            \item {\bf Counting cliques in each row.}
            It remains to compute $N_{r}$ for a given row $r$.
            Suppose an $s$-tuple $v_{1}, \dotsc, v_{s}$ satisfies row $r$.
            Then, for any fixed coordinate $\ell \in [d]$, we have that $v_{1, \ell}, \dotsc, v_{s, \ell}$ satisfy the following.
            For all pairs $i, j$, $v_{i, \ell}, v_{j, \ell}$ satisfy the equality and adjacency conditions of $r$ for $G_{F(i, j), \ell}$ in all decompositions, where $G_{F(i, j), \ell}$ is the $\ell$-th input graph in the factored component assigned to $(h_{i}, h_{j})$.
            We thus call the $s$-tuple $v_{1}, \dotsc, v_{s}$ {\bf $\ell$-relevant}.
            We note that for a given $s$-tuple, $\ell$-relevance can be checked in $O(|\cC| s^{2})$ time, since we check $O(1)$ conditions for at most $s^2$ pairs of vertices and $|\cC|$ decompositions.
            
            Note that for distinct $\ell$, whether $v_{1, \ell}, \dotsc, v_{s, \ell}$ satisfy the equality and adjacency relations are independent.
            Thus, to compute $N_{r}$, we first compute the number of $s$-tuples in coordinate $\ell$ that are $\ell$-relevant for each $\ell$.
            We can then obtain $N_{r}$ by taking the product over all coordinates.

            To count the number of $\ell$-relevant tuples, we note that it suffices to consider $s$-tuples of vertices in $\bigcup_{F, i, j} G_{F(i, j), \ell}$ which is a graph with at most $|\cC| \binom{s}{2} n$ vertices.
            Thus, to enumerate over all $s$-tuples and check each for $\ell$-relevance requires $\bigO{|\cC|^{s} s^{2 s} n^{s} |\cC| s^{2}}$
            The count of $N_{r}$ for a single factored component can therefore by computed in $\bigO{k |\cC|^{S} s^{2 s} n^{s} |\cC| s^{2}}$ time.
        \end{enumerate}
        
        The complexity of the algorithm overall is thus
        \begin{equation*}
            \bigO{2^{|\cC| s^{2} k} s^{2} k + k |\cC|^{s} s^{2 s} n^{s} |\cC| s^{2}} = \bigO{g'(s, k) n^{s}}
        \end{equation*}
        for some function $g'(s, k)$ independent of $n$ since $|\cC| \leq c(s, k)$.
    \end{proof}

    Then, we count $\countH(V[G_{d}])$ for all $d$ and sum in time $O\left( k g'(s, k) n^{s} \right)$.
    This concludes the proof of \Cref{thm:counting-subgraphs} by taking $g(s, k) = k g'(s, k)$.
\end{proof}

\section{Reachability on Factored Graphs}
\label{sec:reachability}

In this section, we show that the parameterized version of the reachability problem on factored graphs is $\xnl$-complete.
Furthermore, we (conditionally) characterize the parameterized time complexity of computing problems in $\xnl$ (in particular, the reachability on factored graphs).
Specifically, we show that all $\xnl$ problems are in $\fpt$ if and only if $\nl \subseteq \dtime\left(n^{C}\right)$ for some constant $C$ that is universal across all problems in $\nl$, establishing a parameterized analog of a major open problem in classical complexity theory.

\subsection{Reachability and Nondeterministic Computation in Logarithmic Space}

Similar to proving \Cref{thm:factored-LFMIS-XP-complete}, we begin with a key lemma that leverages an efficient factorization of the configuration graphs of log-space-bounded nondeterministic Turing machines.
As we will see later, this lemma is central to establishing both the $\xnl$-completeness result and the equivalence result.

\begin{lemma}\label{lem:reachability-key-lemma}
    Let $L$ be a language decidable by a nondeterministic Turing machine using $S \log n$ space for some constant $S$.
    Then, $L$ is mapping reducible using $O(S^2 n^8 \log^4 n)$ time to the reachability problem on factored graphs with parameter $O(S^2)$.
    In particular, the reduction maps every input of length $n$ to a factored graph of complexity $(O(n^4 \log ^2n), O(S^2))$.
\end{lemma}

It is a well-known fact that the standard reachability problem is $\nl$-complete (see e.g.\ \cite{sipser1996introduction}).
We briefly review the standard reduction from an arbitrary language $L \in \nl$ to the reachability problem, and then motivate our proof for \Cref{lem:reachability-key-lemma}.

Given a language $L \in \nl$, there exists a non-deterministic Turing machine $M$ such that $M$ decides $L$ using at most $S \log n$ space on its work tape on any input $x$ of size $n$, where $S$ is a constant.
The standard reduction constructs a \emph{configuration graph} $G$, whose nodes represent configurations of $M$ on $x$. 
For such a construction, each node would have to include information about the current state, the work tape content, and the positions of the two tape heads.
There is a directed edge $(c_1, c_2)$ between two configurations $c_1$ and $c_2$ if and only if $c_2$ is one of the possible next configurations of $M$ on $x$ starting from $c_1$.

Naively, this graph requires up to $\Omega(n^{S})$ vertices to include all the possible contents of the work tape, which has length $S \log n$.
However, this is not sufficient for showing the $\xnl$-hardness of reachability on factored graphs according to \Cref{def:xp-xnl-complete}, as for a language in $\xnl$, we would replace $S$ with some function $f(k)$, but then the reduction would have a running time of at least $\Omega(n^{f(k)})$, where the exponent of $n$ depends on $k$.
Our goal for the proof of \Cref{lem:reachability-key-lemma} is to build $G$ as a factored graph of complexity $(n^{O(1)}, g(S))$ for some function $g$. 
In particular, this means that we want to ``hide'' the exponent $S$ into the number of input graphs in the factored representation. 
This can be achieved by dividing the work tape into $S$ segments, each of which has length $\log n$\footnote{We assume that both $S$ and $\log n$ are positive integers without loss of generality.}. 
In such a case, the length of a segment is $\log n$ and the number of possible configurations on that segment is a small fixed polynomial of $n$ independent of $S$.
On a high level, for each segment we build a graph factor that represents the set of all sub-configurations on that segment, where a sub-configuration encodes the contents of the work tape on the given segment. 
If the work tape head currently lies in a segment (known as the \actS{} segment), the vertex furthermore encodes the state and tape head positions of the Turing machine.
In the end, we use a tensor product to combine all the graph factors and recover the set of all configurations of $M$ on $x$.
Since the transition of the Turing machine is determined locally (by the current state and values under the tape heads), the factored graph correctly encodes the configuration graph of $M$ on $x$.
We now proceed to the formal proof of the key lemma.

\begin{proof}[Proof of \Cref{lem:reachability-key-lemma}]
    Let $L$ be a language decidable by a nondeterministic Turing machine $M$ using $S \log n$ space for some constant $S$.
    Without loss of generality, we assume that when $M$ reaches the accept or reject state, it overwrites the entire work tape with $\perp$ and moves the work tape head to the left-most entry in the corresponding halting state.
    We also assume that both $S$ and $\log n$ are integers (otherwise replace them with $\lceil S \rceil$ and $\lceil \log n \rceil$, respectively).
    Given some input $x$ of size $n$, a configuration of $M$ on $x$ is a setting of the state, the work tape, and the position of the two tape heads.
    We define a \emph{segmented configuration}.

    \begin{definition}
        \label{def:segmented-configuration}
        Let $c$ be a configuration of $M$ on $x$.
        Let $i \in [S]$.
        A \emph{configuration of $M$ on $x$ on segment $i$} consists of:
        \begin{enumerate}
            \item a status: \actS{} or \inactS{};
            \item the state if status is \actS{} and $\nullS$ otherwise;
            \item the positions of the input and work tape head if status is \actS{} and $\nullI, \nullW$ otherwise;
            \item the work tape between locations $(i - 1)\log n + 1$ and $i \log n$.
        \end{enumerate}
        A \emph{segmented configuration of $c$} consists of $(c_{1}, \dotsc, c_{S})$ where $c_{i}$ is a configuration of $M$ on $x$ on segment $i$ and the following are satisfied:
        \begin{enumerate}
            \item the work tape of $c_{i}$ agrees with the work tape of $c$ between locations $(i - 1)\log n + 1$ and $i \log n$.
            \item $c_{i}$ is \actS{} if and only if the work tape head of $c$ is between locations $(i - 1)\log n + 1$ and $i \log n$.
            \item the state of the \actS{} $c_{i}$ agrees with $c$.
            \item the positions of the two tape heads of the \actS{} $c_{i}$ agrees with $c$.
        \end{enumerate}
        In this case, we say that $c_i$ is a \emph{sub-configuration} of $c$ on segment $i$.
    \end{definition}

    \paragraph*{Factored Graph Construction} 
    We will define a factored graph $G$ whose nodes are segmented configurations. 

    \begin{enumerate}
        \item {\bf Inactive Segment Graphs}
        For each $i \in [S]$, define $I_{i}$ to be the graph with vertices corresponding to the set of all configurations of $M$ on $x$ on segment $i$ where the status is \inactS{} (and therefore state $\nullS$ and tape head positions $\nullI, \nullW$).
        The edge set of $I_i$ contains a self-loop on every vertex.

        \item {\bf Active Segment Graphs}
        For each $i \in [S]$, define $A_{i}$ to be the graph with vertices corresponding to the set of all configurations of $M$ on $x$ on segment $i$ where the status is \actS{}. For vertices $v_1, v_2$ in $A_i$, there is an edge $(v_1, v_2)$ if and only if there exist configurations $c_1, c_2$ of $M$ on $x$ such that
        \begin{itemize}
            \item $c_2$ is one of the possible next configurations of $M$ on $x$ starting from $c_1$, and
            \item $v_1, v_2$ are sub-configurations of $c_1, c_2$ on segment $i$, respectively.
        \end{itemize}
        In particular, this means that segment $i$ remains active after transitioning from $c_1$ to $c_2$.      
        
        \item {\bf Transition Segment Graphs}
        For each $2 \leq i \leq S$, define $T_{i-1, i}$ to be the graph with vertices $V(A_{i - 1} \times I_{i}) \cup V(I_{i - 1} \times A_{i})$ corresponding to the set of all pairs of configurations $(v_{i-1}, v_i)$ of $M$ on $x$ on segment $i-1$ and $i$ such that exactly one of them is \actS{} (and so the other is \inactS{}). 
        In $T_{i-1, i}$, there is an edge $((v_{i-1}, v_i), (u_{i-1}, u_i))$ if and only if there exist configurations $c_1, c_2$ of $M$ on $x$ such that
        \begin{itemize}
            \item $c_2$ is one of the possible next configurations of $M$ on $x$ starting from $c_1$, and
            \item $v_{i-1}, v_i$ are sub-configurations of $c_1$ on segments $i-1, i$ and $u_{i-1}, u_i$ are sub-configurations of $c_2$ on segments $i-1, i$, and
            \item the status on both segments $i-1$ and $i$ flip from $c_1$ to $c_2$
        \end{itemize}
        In other words, the work tape head moves from one segment to the other after the transition.
    \end{enumerate}

    Now, we define the following collection of tensor products
    \[
    \sconfig := \left\{ \begin{array}{c}
         A_1 \times I_2  \times \dotsb \times I_{S-1} \times I_S,\\
         I_1 \times A_2  \times \dotsb \times I_{S-1} \times I_S,\\
         \vdots \\
         I_1 \times I_2  \times \dotsb \times I_{S-1} \times A_S
    \end{array}\right\} \cup
    \left\{ \begin{array}{c}
         T_{1, 2} \times I_3  \times \dotsb \times I_S,\\
         I_1 \times T_{2, 3}  \times \dotsb  \times I_S,\\
         \vdots \\
         I_1 \times \dotsb \times I_{S-2} \times T_{S-1, S}
    \end{array}\right\}
    \]

    Then, we simply define $G$ as the union of all the tensor products in $\sconfig$
    \[
    G := \bigcup_{G' \in \sconfig} G'.
    \]

    Observe that the vertex set of $G$ is
    \begin{equation*}
        V(G) = \bigcup_{i = 1}^{S} V(I_{1} \times \dotsb \times I_{i - 1} \times A_{i} \times I_{i + 1} \times \dotsb \times I_{S})
    \end{equation*}
    since the tensor product is associative.
    
    \paragraph*{Reachability Computes Turing Machine Acceptance}
    It suffices to show that $G$ is the configuration graph of $M$ where nodes correspond to configurations of $M$ on input $x$ and edges encode valid transitions of $M$.

    First, we show that the nodes of $G$ are exactly the segmented configurations of $M$.
    Consider a node in $G$.
    Suppose the node is in $I_{1} \times \dotsb \times I_{i - 1} \times A_{i} \times I_{i + 1} \times \dotsb \times I_{S}$ for some $i \in [S]$.
    This node determines the configuration of $M$ on $w$ whose state and tape head positions are given by the vertex in $A_{i}$ and whose work tape is given by the concatenation of all $S$ components.
    We note here that exactly one segment contains the tape head in each configuration, so that in each vertex exactly one coordinate is \actS{}.
    
    Conversely, consider a configuration of $M$ on $x$.
    The tape head $i_{W}$ lies in some segment $1 \leq i \leq S$.
    Thus, there is a node in $I_{1} \times \dotsb \times I_{i - 1} \times A_{i} \times I_{i + 1} \times \dotsb \times I_{S}$ corresponding to the segmented configuration.

    Next, we argue that two nodes in $G$ are connected if and only if the corresponding configurations have a valid transition between them.
    Suppose $(c_1, c_2)$ is a valid transition for $M$ on $x$.
    Note that the work tape head position in $c_1, c_2$ differs by at most one, and must either be in the same segment or adjacent segments.
    In the former case, the corresponding vertices form an edge given by $I_{1} \times \dotsb \times I_{i - 1} \times A_{i} \times I_{i + 1} \times \dotsb \times I_{S}$.
    In the latter case, the corresponding vertices form an edge given by $I_{1} \times \dotsb \times I_{i - 2} \times T_{i - 1, i} \times I_{i + 1} \times \dotsb \times I_{S}$.

    On the other hand, suppose there is an edge between nodes $(u, v)$ in $G$.
    Denote $u = (u_1, \dotsc, u_{S})$ and $v = (v_1, \dotsc, v_{S})$.
    This edge must either come from $I_{1} \times \dotsb \times I_{i - 1} \times A_{i} \times I_{i + 1} \times \dotsb \times I_{S}$ or $I_{1} \times \dotsb \times I_{i - 2} \times T_{i - 1, i} \times I_{i + 1} \times \dotsb \times I_{S}$ for some $i$.
    
    We begin with the former case.
    Then, since $I_{i}$ contains only self loops, we have $u_{j} = v_{j}$ for all $j \neq i$ and $(u_{i}, v_{i}) \in E(A_{i})$.
    Thus, there are some configurations $c_{1}, c_{2}$ for which there is a valid transition from $c_{1}$ to $c_{2}$ for $M$ on $x$ and $u_{i}, v_{i}$ are sub-configurations of $c_{1}, c_{2}$ on segment $i$.
    While $c_{1}, c_{2}$ may not correspond to $u, v$, we note that the validity of a transition is determined only by the current state, the position of the tape heads, and the value of the tape under the two heads.
    Thus, if there is a valid transition from $c_{1}$ to $c_{2}$, there must also be a valid transition from the $u$ to $v$ since $u, c_{1}$ agree on the state, tape head positions, and the value of the tape under the two heads (since they agree on the $i$-th segment of the work tape) and $v, c_{2}$ agree on the relevant values as well.
    
    To tackle the latter case, assume without loss of generality that $u_{i - 1} \in V(A_{i - 1})$ and $v_{i} \in V(A_{i})$.
    Again, we have $u_{j} = v_{j}$ for all $j \not\in \set{i - 1, i}$.
    Then, since $(u_{i - 1}, u_{i}), (v_{i - 1}, v_{i})$ are adjacent in $T_{i - 1, i}$, there are configurations $c_{1}, c_{2}$ such that $u_{i - 1}, u_{i}$ are sub-configurations of $c_{1}$ and $v_{i - 1}, v_{i}$ are sub-configurations of $c_{2}$ on segments $i - 1, i$.
    Again, while $c_{1}, c_{2}$ may not correspond to $u, v$, we note that there is a valid transition from $c_{1}$ to $c_{2}$ if and only if there is one from $u$ to $v$ using similar arguments as in the first case.

    We have thus shown that $G$ has a vertex set in one-to-one corresponding to the configurations of $M$ on $x$ and vertices are adjacent if and only if there is a valid transition between the corresponding configurations.

    \paragraph*{Reduction Complexity}
    As defined in \Cref{def:segmented-configuration}, the number of total possible configurations of $M$ on input $x$ on segment $i$ is at most $O(n^2 \log n)$. 
    The vertex sets of $I_i$ and $A_i$ are subsets of all possible configurations of $M$ on $x$ on segment $i$ and the vertex set of $T_{i-1, i}$ is a subset of pairs of such configurations.
    Therefore, the size of each graph factor is bounded above by $O(n^4 \log^2 n)$. 
    Moreover, it follows from the definition of $\mathcal F$ that the number of input graph factors is $O(S^2)$.
    This shows that $G$ is a factored graph of complexity $(O(n^4 \log^2 n), O(S^2))$.
    Since generating each input graph factor requires at most quadratic time in its size, the total time complexity of the reduction is $O(S^2 n^8 \log^4 n)$, which completes the proof of \Cref{lem:reachability-key-lemma}.
    \end{proof}

    Using the key \Cref{lem:reachability-key-lemma}, we can now prove \Cref{thm:reachability-xnl-complete} without too much difficulty.
    We begin with the proof of the $\xnl$-completeness result in the next subsection.

    \subsection{\textbf{XNL}-completeness of Reachability on Factored Graphs}
    In this section, we show that the parameterized version of reachability on factored graphs is $\xnl$-complete.

    \begin{theorem}[Part 1 of \Cref{thm:reachability-xnl-complete}]\label{thm:reachability-part-1}
        Reachability on factored graphs is $\xnl$-complete under $\fpt$-reductions.
    \end{theorem}
    \begin{proof}
        We first show that reachability on factored graphs is in $\xnl$, and then prove its $\xnl$-hardness using the key \Cref{lem:reachability-key-lemma}.

        \paragraph*{Membership in XNL}
        Let $G = f(G_1, \dotsc, G_k)$ be a factored graph of complexity $(n, k)$ and let $s, t$ be the specified source and target vertices.
        Each vertex in $G$ can be specified by $O(k \log n)$ bits, since each vertex is a tuple of at most $k$ coordinates and each coordinate is a vertex in an input graph $G_i$ with at most $n$ vertices.
        We define the following non-deterministic Turing Machine which maintains on its tape 1) the current vertex on the path, 2) the next vertex on the path, and 3) the number of vertices on the path considered so far.
        The machine begins by writing $s$ as the current vertex on the path and initializes the counter to $0$.
        Then, the machine repeatedly guesses the next vertex on the path non-deterministically, and checks whether the two vertices are adjacent in $G$ by verifying $O(k)$ equality or adjacency conditions as in the proof of \Cref{thm:counting-subgraphs}.
        If the sequence of at most $n^{k}$ vertices computes a path to $t$, the machine accepts.
        On the other hand, if the counter reaches $n^{k} + 1$ or some adjacency relation does not hold, the machine rejects.
        Note that the space consumed is $O(k \log n)$, which proves membership in $\xnl$.

        \paragraph*{XNL-hardness under FPT-reduction}
        Let $L$ be a parameterized language in $\xnl$.
        Then, there exists a nondeterministic Turing machine $M$, such that for each parameter $k \in \N$, $M$ decides $L$ using $O(f(k) \log n)$ space on inputs of length $n$, where $f$ is a function of $k$.
        By applying the mapping used in the reduction of \Cref{lem:reachability-key-lemma}, we obtain a mapping reduction from $L_k$ to the reachability problem on factored graphs with parameter $O(f(k)^2)$ using $O(f(k)^2 n^{O(1)})$ time, which proves $\xnl$-hardness.
    \end{proof}

    \subsection{Equivalence to the Open Problem}
    In this section, we complete the proof of \Cref{thm:reachability-xnl-complete} by establishing the remaining equivalence result:
    
    \begin{theorem}[Part 2 of \Cref{thm:reachability-xnl-complete}]\label{thm:reachability-xnl-part-2}
        $\xnl \subseteq \fpt$ if and only if $\nl \subseteq \dtime(n^C)$ for some absolute constant $C$.
    \end{theorem}

    It turns out that this proof uses the same major building blocks as the previous proof in a slightly different way.

    \begin{proof}[Proof of \Cref{thm:reachability-xnl-part-2}]
        For the forward direction, suppose $\xnl \subseteq \fpt$.
        Then in particular, the reachability problem on factored graphs is in $\fpt$.
        Now, consider any language $L \in \nl$, which is decidable by some nondeterministic Turing machine $M$ using $S \log n$ space for some constant $S$.
        By \Cref{lem:reachability-key-lemma}, given an input $x$ of length $n$, we construct an instance of the reachability problem on a factored graph $G$ of complexity $(O(n^4 \log^2 n), O(S^2))$.
        This construction takes $O(S^2 n^{8} \log^4 n)$ time, and reachability on $G$ simulates the computation of $M$ on $x$.
        But since reachability on factored graphs is in $\fpt$, it follows that reachability on $G$ can be solved in time $O(f(S^2) n^{4C_0} \log^{2C_0} n)$ for some function $f$ and constant $C_0$.
        It is important to note here that the constant $C_0$ comes from the fixed-parameter tractability assumption of reachability on factored graphs, which means $C_0$ does not depend on $S$ or the specific language $L \in \nl$.
        Combining the mapping reduction and the $\fpt$ algorithm for reachability, we obtain an algorithm for $L$ that runs in time $O(n^{4C_0} \log^{2C_0} n + n^8 \log^{4} n)$.
        The forward direction then follows by choosing $C := 4C_0 + 9$.
        
        Conversely, suppose $\nl \subseteq \dtime(n^C)$ for some absolute constant $C$.
        To show that $\xnl \subseteq \fpt$, it now suffices to show that reachability on factored graphs is in $\fpt$ due to its $\xnl$-completeness under $\fpt$-reductions.
        Note that the $\xnl$ algorithm used in the proof of \Cref{thm:reachability-part-1} is in fact an $\nl$ algorithm for each fixed parameter $k$.
        Thus, if $\nl \subseteq \dtime \left( n^{C} \right)$ for some $C$ independent of $k$, then for any fixed $k$, reachability on factored graphs of complexity $(n, k)$ can be computed in time $O(k^{C} n^{2 C})$ (as factored graphs of this complexity have input size $O(k n^2)$) and is therefore in $\fpt$.
    \end{proof}

\subsection{Fixed-parameter Tractability under Single Graph Operation}
To conclude the section on reachability, we note that the difficulty of computing reachability on factored graphs crucially lies in the fact that the factored graphs use multiple graph operations.
In particular, while our lower bound construction primarily uses the Cartesian product to encode vector edges, the following lemma shows that reachability on factored graphs using Cartesian product or union alone is in $\fpt$.

\begin{proposition}
    \label{prop:reachability-one-operation}
    Suppose $G = G_1 \circ G_2 \circ \dotsb \circ G_k$ for $\circ \in \set{\cart, \cup}$.
    Then, for any pair of vertices $s, t \in G$, reachability can be computed in $O(k n^{2})$ time.
    In particular, reachability is in fixed-parameter tractable.
\end{proposition}

\begin{proof}
    Suppose $G = G_1 \cart \dotsb \cart G_k$.
    Then, $s = (s_1, \dotsc, s_{k})$ and $t = (t_1, \dotsc, t_{k})$.
    We claim that $s$ reaches $t$ if and only if $s_{i}$ reaches $t_{i}$ for all $i \in [k]$.
    Suppose $s$ reaches $t$.
    Denote the path $P = (s = v_{0}, v_{1}, \dotsc, v_{\ell} = t)$.
    For each $i$, consider the sub-path $P_{i}$ of edges where the $i$-th coordinate changes.
    Each $P_{i}$ gives an edge from $s_{i}$ to $t_{i}$.
    Conversely, each $s_{i}$ reaches $t_{i}$, we can construct a path $P$ first from $s$ to $(t_1, s_2, \dotsc, s_{k})$ and repeat the process for each coordinate to construct a path from $s$ to $t$.
    Thus, it suffices to compute reachability on each $G_{i}$, requiring total time $O(k n^{2})$.
    
    Suppose $G = G_1 \cup \dotsb \cup G_m$.
    Note that $G$ can be constructed in $O(k n^{2})$ time, so that we can construct $G$ and solve reachability explicitly in $O(k n^{2})$ time.
\end{proof}

\section{Conclusion and Future Work}
\label{sec:conclusion}

We have studied the computational complexity of various problems on factored graphs.
Even among problems with polynomial time algorithms on explicit graphs, we have shown that their parameterized complexity when the input is represented as a factored graph can differ quite drastically.
In the context of parameterized complexity, we have sought to characterize which problems on factored graphs are in $\fpt$.
On the positive side, counting the number of copies of a small clique is in $\fpt$.
On the negative side, LFMIS is \emph{unconditionally} not in $\fpt$.
Finally, we show that determining whether reachability is in $\fpt$ is equivalent to determining whether $\nl \subseteq \dtime(n^{O(1)})$.

Can the unconditional lower bounds for LFMIS on factored graphs be used to prove similar unconditional lower bounds for other parameterized problems?  One obstacle to doing this is the gap in complexity between the $\ptime$-complete LFMIS problem and the easily parallelizable problems that form the bulk of the literature in parameterized complexity. However, reductions in fine-grained complexity often cut across traditional complexity classes, e.g., \cite{kunnemann2017dp}. So we do not know a reason why this should not also be the case here. Either finding such unconditional results or explaining their impossibility would both be interesting.  We could also hope to prove similar results for other $\ptime$-complete problems.  

While our lower bounds separate the problems of interest from $\fpt$, they do not rule out significant improvements on the naive $n^{O(k)}$ algorithm of computing the factored graph $G$ explicitly and solving the problem on $G$ itself.
For example, our LFMIS lower bound only unconditionally rules out algorithms with time $n^{o(\sqrt{k})}$.
An interesting open problem is to provide a more \emph{fine-grained} analysis of the complexity of factored graph problems, possibly distinguishing between the number of product and union operations.  

In this work, we have chosen to study factored graphs under graph products and unions, specifically the Cartesian and tensor products.
A natural extension is to consider other products and operations on graphs, or other interesting objects. Factored problems on bit strings were introduced in \cite{dalirrooyfard2020factor}, and implicitly on integer-valued vectors in \cite{an2022fine}.  Do similar results hold for factored problems for these input domains and operations?

\section*{Acknowledgements}
We would like to thank Antonina Kolokolova, Anthony Ostuni, and anonymous reviewers for their many helpful comments and suggestions.

\bibliographystyle{alpha}
\bibliography{references}

\end{document}